\pdfoutput=1
\documentclass[singlespacing]{elsart}
\usepackage{icarus}
\usepackage{graphicx}
\usepackage{natbib}
\bibpunct{(}{)}{;}{a}{,}{,}
\usepackage{times}
\usepackage{fancyhdr}
\normalsize

\begin{document}

\textsc{\it Icarus, in press}\\[0.5cm]

%\begin{frontmatter}

\title{Generation of equatorial jets by large-scale latent heating on the giant planets}
\author{Yuan Lian and}
\author{Adam P. Showman}

\address{Lunar and Planetary Laboratory, the University of
                        Arizona, Tucson, AZ, 85721 USA}

\textsc{\small Accepted by Icarus, in press}\\[0.5cm]
%\date{Accepted by Icarus, in press}

\maketitle

\begin{abstract}

Three-dimensional numerical simulations show that large-scale latent heating
resulting from condensation of water vapor can 
produce multiple zonal jets similar to those on the gas giants 
(Jupiter and Saturn) and ice giants (Uranus and Neptune).  For
plausible water abundances (3--5 times solar on Jupiter/Saturn and
30 times solar on Uranus/Neptune), our simulations produce 
$\sim20$ zonal jets for Jupiter and Saturn and 3 zonal jets on Uranus
and Neptune, 
similar to the number of jets observed on these planets.
Moreover, these Jupiter/Saturn cases produce equatorial {\it superrotation}
whereas the Uranus/Neptune cases produce equatorial {\it subrotation},
consistent with the observed equatorial jet direction on these planets.
Sensitivity tests show that water abundance, planetary rotation rate,
and planetary radius are all controlling factors, with water playing
the most important role; modest water abundances, large planetary radii,
and fast rotation rates favor equatorial superrotation,
whereas large water abundances favor equatorial subrotation regardless
of the planetary radius and rotation rate.   Given the larger radii,
faster rotation rates, and probable lower water abundances of Jupiter 
and Saturn relative to Uranus and Neptune,
our simulations therefore provide a possible mechanism for the existence of
equatorial superrotation on Jupiter and Saturn and the lack of superrotation
on Uranus and Neptune. Nevertheless, Saturn poses a possible difficulty,
as our simulations were unable to explain the unusually high speed
($\sim400\rm\,m\,s^{-1}$) of that planet's superrotating jet.
The zonal jets in our simulations exhibit 
modest violations of the barotropic and Charney-Stern stability criteria. 
Overall, our simulations, while idealized, support the
idea that latent heating plays an important role in generating
the jets on the giant planets.

\end{abstract}


\section{Introduction}

The question of what causes the prominent east-west (zonal) jet
streams and banded cloud patterns on Jupiter, Saturn, Uranus, and 
Neptune remains a major unsolved problem in planetary science.
On Jupiter and Saturn, there exist $\sim20$--30 jets, including
a broad, fast superrotating (eastward) equatorial jet. In contrast,
Uranus and Neptune exhibit only $\sim3$ jets each, with high-latitude
eastward jets and a subrotating (westward) equatorial jet.
All four planets also exhibit a variety of compact vortices, waves, 
turbulent filamentary regions, short-lived convective events,
and other local features.  
Recent observational studies demonstrate that small eddies pump momentum 
up-gradient into the zonal jets on Jupiter and Saturn 
\citep{2006Icar..185..430S, Del_Genio2007}, which strongly suggests
that cloud-layer processes are important in jet formation, although
this does not exclude a possible role for the deep interior too.
Scenarios to explain these diverse observations range from the 
``shallow-forcing'' scenario, in which jet generation occurs via
injection of turbulence, absorption of solar 
radiation and latent heat release in the cloud layer, to the 
``deep-forcing'' scenario, in which the jet formation results
from convection occurring throughout the molecular envelope
\citep[for reviews see][]{ingersoll-etal-2004, 2005RPPh...68.1935V}.   
Over the past several decades, a variety of idealized models
have been developed that successfully produce banded zonal flows
reminiscent of those on the giant planets 
\citep{1978JAtS...35.1399W, cho-polvani-1996a, 1998JAtS...55..611H,
Heimpel2007, lian-showman-2008}.

Despite these successes, the equatorial jet direction and magnitude
 have proved to be formidable puzzles that are difficult to explain.
Many published models predict the same equatorial jet direction for
all four giant planets and thereby fail to provide a coherent
explanation that encompasses both the gas giants (Jupiter/Saturn)
and the ice giants (Uranus/Neptune).  Under relevant conditions, 
one-layer shallow-water-type 
models generally produce westward equatorial flow for all four
planets \citep{1996PhFl....8.1531C,1999PhFl...11.1272I, showman-2007,
scott-polvani-2007}, consistent with Uranus and Neptune but not 
Jupiter and Saturn.  In the parameter regime of giant planets,
some shallow-water models have produced equatorial superrotation 
\citep{scott-polvani-2008}, but as yet these models make no predictions
for why superrotation should occur on Jupiter and Saturn but not Uranus
and Neptune.  Some
recent three-dimensional (3D) shallow-atmosphere models can also produce 
equatorial superrotation under specific conditions 
\citep{Williams_2006,Williams_2002, Williams_2003_3, Williams_2003_2, 
Williams_2003_1, 2005P&SS...53..508Y, lian-showman-2008}, but
this has generally required the addition of {\it ad hoc} forcing,
and even if such forcing were plausible, it is unclear why it
would occur on Jupiter/Saturn but not Uranus/Neptune. 
In contrast, the deep convection models 
produce equatorial superrotation in most cases \citep{2001GeoRL..28.2557A, 
2001GeoRL..28.2553C, 2005Natur.438..193H,Heimpel2007, glatzmaier-etal-2008}, 
consistent with Jupiter and Saturn but inconsistent with Uranus and Neptune.
In the context of deep convection models, 
\citet{aurnou-etal-2007} proposed that 
Uranus and Neptune are in a regime where geostrophy breaks down
in the interior, leading to turbulent mixing of angular momentum
and a westward equatorial jet.  However, this mechanism 
occurs only at heat fluxes greatly exceeding than those observed on
Uranus and Neptune \citep{aurnou-etal-2007}.  Moreover,
given that the heat fluxes on Jupiter and Saturn exceed those on Uranus 
and Neptune, one might expect the mechanism to apply more readily to 
the former pair than the latter pair; if so, one should see equatorial 
subrotation on Jupiter/Saturn yet superrotation on Uranus/Neptune, 
backward from the observed equatorial jet directions.  
\citet{schneider-liu-2009} developed a 3D numerical model that produced banded 
zonal jets and an equatorial superrotation on Jupiter.  This is the first
model that combines the deep convection (via a simple convective adjustment
scheme) and absorption of solar radiation in a 
shallow atmosphere.  However, their model extends to only 3 bars pressure
and thus neglects the effects of latent heating, which may be crucial
in generating horizontal temperature contrasts in the cloud layer.
Furthermore, it is unclear whether their model can produce equatorial 
subrotation on Uranus and Neptune by the same mechanism.  It is fair to say 
that we presently lack a coherent explanation for the equatorial jets
that encompasses both the gas giants (Jupiter/Saturn) and the ice 
giants (Uranus/Neptune).

Here, we test the hypothesis that large-scale latent heating associated with
condensation of water vapor can pump the zonal jets on the four giant planets.
This hypothesis has been repeatedly suggested over the past 40 years
\citep{barcilon-gierasch-1970, 1976Icar...29..445G, 2000Natur.403..630I, 
2000Natur.403..628G}, but this idea has not yet been adequately
tested in numerical models.  While observations cannot yet constrain the 
existence of large-scale latent heating, abundant evidence nevertheless
exists for moist convection on the giant planets.
Lightning has been identified in nightside
images of Jupiter from Voyager, Galileo, Cassini, and New Horizons; 
these flashes typically occur within localized, opaque clouds that grow 
to diameters up to $\sim3000\,$km over a few days, indicating
convective activity.  The lightning illuminates finite regions
on the cloud deck, indicating that the flashes occur at depths 
of 5--10 bars, in the expected water condensation region 
\citep{borucki-williams-1986, dyudina-etal-2002}.
Near such storms, clouds are
sometimes observed whose tops are deeper than 4 bars, where the
only condensate is water \citep{1998Icar..135..230B, 2000Natur.403..628G}. 
On Saturn, electrostatic discharges
presumably caused by lightning have been identified, as have explosive
convective clouds that probably cause them \citep{Porco2005,
dyudina-etal-2007}.
Whistlers and electrostatic discharges indicating the presence of 
lightning have also been detected on Uranus and Neptune 
\citep{zarka-pedersen-1986, gurnett-etal-1990,
kaiser-etal-1991}, suggesting that moist convection occurs on
these planets too.  For plausible water abundances (a few
times solar or greater), latent heating can cause local temperature
increases great enough to have important meteorological effects.

To date, numerical models of jet formation have generally
not included moisture and its latent heat release. 
Most two-dimensional (2D) and shallow-water models 
adopt  forcing that injects turbulence everywhere simultaneously
and is confined to a small range of wavenumbers \citep[e.g.][]
{1998JAtS...55..611H, scott-polvani-2007}, which does not
capture the sporadic and localized nature of moist-convective
events.   Likewise, existing 3D models of Jovian jet formation
have been dry (no water vapor) and force the flow by imposing latitudinal
temperature differences rather than including moist convection 
\citep[e.g.][]{Williams_2006,Williams_2002, Williams_2003_3, Williams_2003_2, 
Williams_2003_1, 2005P&SS...53..508Y, lian-showman-2008}.
Notable efforts in the right direction are the one-layer studies by
\citet{Li-2006} and \citet{showman-2007}, which adopted a forcing
explicitly intended to represent the effects of moist convection. 
\citet{Li-2006} adopted a quasigeostrophic model and
introduced isolated vorticity patches to represent moist-convective storms; 
\citet{showman-2007} adopted the shallow-water
equations and introduced isolated mass pulses to represent the
moist convection.   These studies show that, under planetary rotation, 
the small-scale turbulent flow can inverse cascade to form large scale 
dynamics:  zonal jets dominate at low latitudes and vortices dominate 
at high latitudes.   Nevertheless,  these models do not explicitly 
include water vapor, and the moist convection events are, rather than 
occurring naturally, injected by hand with prescribed sizes, lifetimes 
and amplitudes.   Studies have also been carried out that
investigate the effects of sophisticated cloud microphysics schemes
on the vertical structure in 1D column models \citep{Del_Genio1990} and 2D 
height/latitude models \citep{nakajima-etal-2000, 2008Icar..194..303P}, but these studies
do not address whether moist convection can pump the jets.

Our previous 3-D studies with imposed latitudinal temperature variation
can produce baroclinic eddies that drive the zonal jets through
inverse-cascade of turbulence \citep{lian-showman-2008}. Those simulations
successfully reproduced some major dynamic features on Jupiter such as
banded zonal winds and equatorial superrotation; they also predicted that the
jets on Jupiter could extend significantly deeper than the eddy accelerations
that pump them.   However, the nature of the imposed forcing
schemes was only a crude parameterization of the processes that produce latitudinal
temperature contrasts.

Here we present three-dimensional (3D) global numerical simulations
using the MITgcm to investigate whether large-scale latent heating can drive the
zonal jets on Jupiter, Saturn, Uranus, and Neptune.  Specifically, we
investigate whether we can explain (i) the approximate number
and speed of jets, and (ii) the direction of the equatorial jet
on all four planets in the context of a single mechanism. We explicitly 
include water vapor as a tracer in our numerical model. 
Section 2 describes the numerical model, 
section 3 presents the simulation results, and section 4 concludes.

\section{Models}

\subsection{Model setup}

We use a global circulation model, the MITgcm, to solve the 3D
hydrostatic primitive equations in pressure coordinates on a sphere.
Previous studies of jet formation on the giant planets have adopted
dry models \citep{lian-showman-2008, Williams_2003_1, showman-2007,
scott-polvani-2007, Li-2006}, but here 
we explicitly treat the transport and condensation of water vapor.
Condensation occurs whenever the relative humidity exceeds 100\%,
and the resultant latent heating is explicitly added 
to the energy equation.  

The system is governed by the horizontal momentum, hydrostatic
equilibrium, mass continuity, energy, and water-vapor equations as follows:

\begin{equation}{{d {\bf v}}\over{dt}}+f\hat{k}\times{\bf v}+\nabla_p\Phi=0
\label{mom}
\end{equation}
\begin{equation}{\partial{{\Phi}}\over{\partial{p}}}=-{1\over\rho}
\label{geopotential}
\end{equation}
\begin{equation}\nabla_p\cdot{\bf v}+{\partial{\omega}\over{\partial{p}}}=0
\label{continuity}
\end{equation}
\begin{equation}
{d\theta \over {dt}}=Q_\theta+{L\over c_p}{\theta\over T}(\delta{ {q-q_s}\over{\tau_s}})
\label{LatentHeat}
\end{equation}
\begin{equation}
{dq\over{dt}}=-{{q-q_s}\over\tau_s}\delta + Q_{\rm deep}
\label{Condensation}
\end{equation}
where ${\bf v}$ is the horizontal wind vector (comprised of zonal
wind $u$ and meridional wind $v$), $\omega=dp/dt$ is vertical wind
in pressure coordinates, $f=2\Omega \sin\phi$ is the Coriolis
parameter (where $\phi$ is latitude and $\Omega$ is the rotation rate
of the planet), $\Phi$ is geopotential, $\hat{k}$ is the
unit vector in the vertical direction (positive upward), $\rho$
is density, $\nabla_p$ is the horizontal gradient operator at a
given pressure level, $d/dt$ is the total derivative operator given by
$d/dt=\partial/\partial t + {\bf v}\cdot\nabla_p + \omega\partial/\partial
p$, $q$ is the water-vapor mixing ratio (defined as kilograms
of water vapor per kilogram of dry H$_2$ air),
and  $\theta=T(p_0/p)^{\kappa}$ is potential temperature. Here $T$
is temperature and $\kappa\equiv R/c_p$, which is a specified constant,
is the ratio of the gas constant to the specific heat at
constant pressure.  In the equations above, density $\rho$ is calculated 
at a given temperature and pressure from the ideal gas law, $\rho = \frac{pm}{R_u T}$, 
where $R_u$ is the universal gas constant and $m$ is the mean molecular mass of 
the moist air, given by 
$m= \frac{m_{H_2}}{1 + q/\epsilon} + q \frac{m_{H_2}}{1 + q/\epsilon}$, where 
$\epsilon = \frac{m_{H_2O}}{m_{H_2}}$ is the ratio of the mass of a water molecule 
to the mass of dry air molecule.  Note that we neglect the density perturbation associated 
with condensate mass loading.  Given the density field, geopotential $\Phi$ is then 
calculated by integrating the hydrostatic equation vertically via Eq. (2).
The reference pressure $p_0$ is taken as $1\,$bar
(note, however, that the dynamics are independent of the choice of 
$p_0$).  Curvature terms are included in ${\bf v}\cdot\nabla{\bf v}$.
In all governing equations Eq.~\ref{mom} -- Eq.~\ref{Condensation}, 
the dependent variables ${\bf v}$, $\omega$, $\Phi$,
$\rho$, $\theta$, and $q$ are functions of longitude $\lambda$, latitude
$\phi$, pressure $p$, and time $t$.

The water-vapor equation (Eq.~\ref{Condensation}) governs the time
evolution of the water-vapor mixing ratio $q$.  There are two source/sink
terms.  The first, $-(q-q_s)\delta/\tau_s$, represents loss through
condensation.  
We apply an ``on-off switch'' $\delta$ to make the condensation events
occur only when the environment is supersaturated: 
when $q > q_s$ then $\delta=1$  and water vapor condenses;
when $q \le q_s$ then $\delta=0$ and water vapor does not condense. 
Here $q_s$ is the saturated water-vapor mixing ratio, given by the
approximate expression
\begin{equation}
q_s={m_{H_2O}\over{m_{H_2}}}{{e_s}\over p},
\label{SpecificHumidity}
\end{equation}
\begin{equation}
e_s={e_0} exp{[-{L\over{R_v}} ({1\over T} -{1\over {T_0}} ) ]},
\label{VaporPressure}
\end{equation}
where $e_s$ is the saturation vapor pressure.
Other constants in Eqs.~\ref{SpecificHumidity}--\ref{VaporPressure}
are given as follows:  
$e_0=609.14 \,\,{\rm Pa}$ is a reference saturation
water-vapor pressure at temperature $T_0=273\,{\rm K}$, 
$R_v=461.0 \rm JK^{-1}kg^{-1}$ is the specific
gas constant of water vapor,  $m_{H_2O}$ is the molecular mass of water and
$m_{H_2}$ is the molecular mass of hydrogen gas. The quantity
$\tau_s$ is the condensation timescale, generally taken as $10^4\,$sec
(almost 3 hours), representative of a typical convective time.
The second term, $Q_{\rm deep}$, represents a source of water
vapor applied near the bottom of the model (see below).

When condensation occurs, we apply the appropriate 
latent heating to the energy equation (second term on right side of
Eq.~\ref{LatentHeat}).  The specific latent heat of condensation is given by
$L=2.5\times 10^6 \rm \, J \, kg^{-1}$.

\citet{2008Icar..194..303P} point out, using simplified one- and
two-dimensional test cases, that cloud microphysics can interact  
with large-scale dynamics on giant planets.  
While recognizing that inclusion of microphysics in 3D is an
important goal for future work, we here make
the simplifying assumption that all of the condensate instantaneously
rains out the bottom of the model.  This allows us neglect cloud
microphysics and thereby sidestep the
numerous complications associated with cloud-particle growth and
settling, the evaporation of falling precipitation, and other microphysical
processes that remain poorly understood --- and whose effects must be
parameterized in large-scale models.  Depending on the complexity of
the adopted schemes, including such processes can introduce potentially 
dozens of new free parameters into the model.  In Earth climate models, 
these parameters are generally tuned using a combination of laboratory
and field data, but it is unclear to what extent such parameter values
(or even the schemes themselves) translate into the giant planet 
context \citep[for discussion see][]{2008Icar..194..303P}.  Given
these difficulties, there is strong merit in exploring the dynamics
in the limiting case without microphysics, as presented here.

True moist convection, which involves the formation of cumulus
clouds and thunderstorms, occurs on length scales much smaller than the
horizontal grid resolutions achievable in most global-scale models (including
ours).  Significant work has gone into developing sub-grid-scale cumulus 
parameterization schemes to represent the effects of this cumulus convection
on the large-scale flow resolved by global models \citep[for reviews see, e.g.,][]
{emanuel-raymond-1993, arakawa-2004}.  Incorporating such a scheme
into a Jovian model is a worthy goal, but such schemes are often complex,
and it is first useful to ascertain the effects of {\it large-scale}
latent heating, that is, latent heating associated with the 
hydrostatically balanced circulation explicitly resolved by the model.  
This is the approach we pursue here.

In general, we expect that the latent heating and decrease in molecular 
mass accompanying condensation/rainout will lead to a vertical structure 
where potential temperature increases with height and molecular mass decreases 
with height. If so, this would imply that condensation would stabilize the 
environment against convection, leading to a virtual potential temperature 
that increases with height.  Given this expectation, we do not include any 
dry convective adjustment scheme in the current simulations. 

To provide a crude representation of evaporating precipitation
and water vapor mixed upward from the deeper atmosphere (below
the bottom of our domain), we apply a source term of water vapor, 
$Q_{\rm deep}$, to the bottom of the model.  This term takes the
form $Q_{\rm deep} = (q_{\rm deep} - q)/\tau_{\rm replenish}$ and is applied
only at pressures exceeding a critical pressure $p_c$, which 
is chosen to be deeper than the deepest possible condensation pressure
for the water-vapor abundance and thermal structure expected
in a given simulation.  Here, $q_{\rm deep}$ is a specified planetary 
water vapor abundance (e.g., 1, 3, 10, or 30 times solar) and 
$\tau_{\rm replenish}$ is a relaxation time.  Our goal is to force
the deep water abundance (at $p>p_c$) to be very close to $q_{\rm deep}$.
The relaxation time, $\tau_{\rm replenish}$, is thus not a free
parameter and is chosen to be very short (typically 5 hours).  This
source term allows
the model to reach a statistical steady state in which the mean
total water vapor content of the atmosphere is nearly constant 
over time --- despite the loss of water via condensation.

\begin{figure}
 \centering
\includegraphics[scale=0.5]{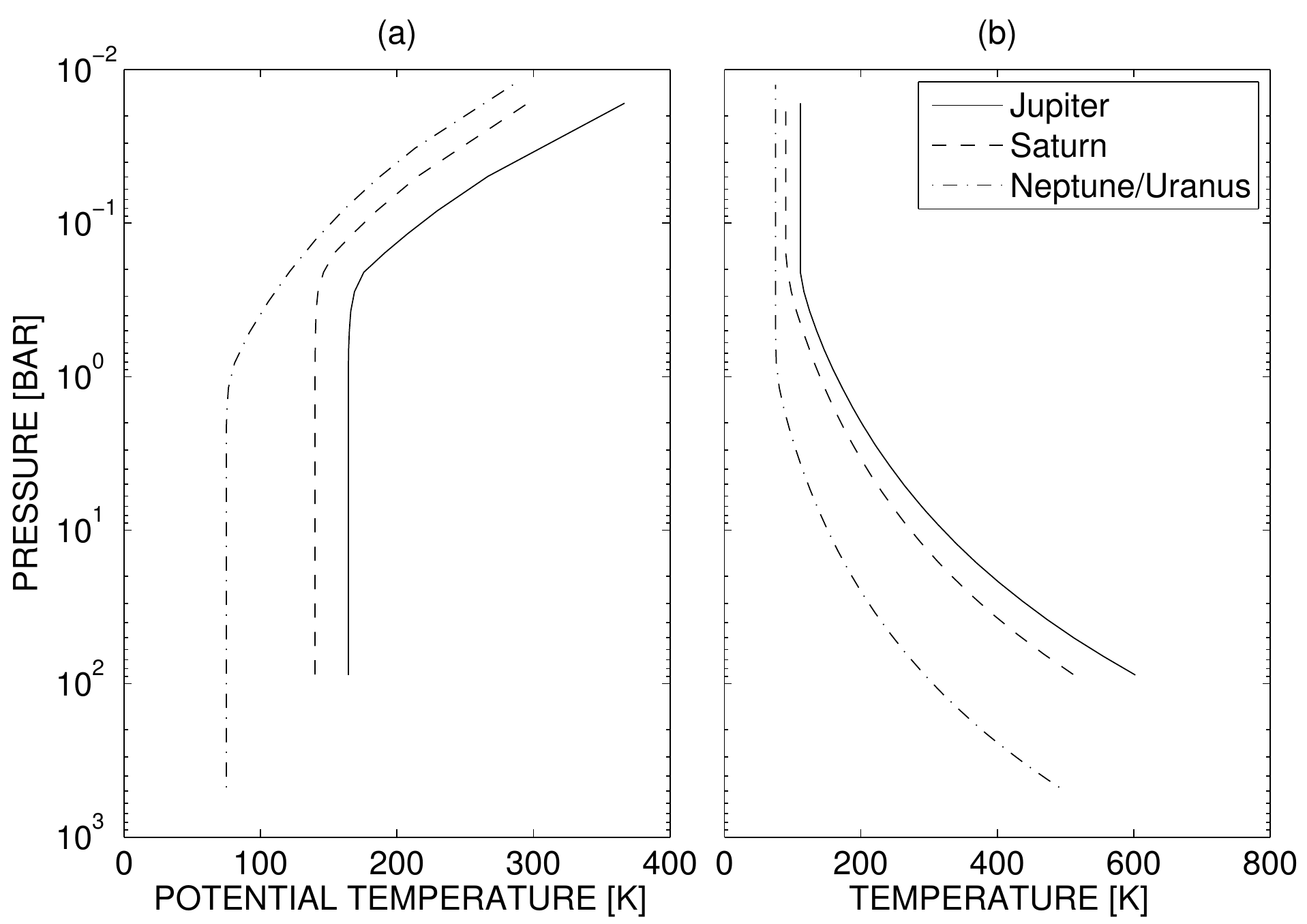}
\caption{Reference temperature-pressure profile for giant planets.}
\label{TP}
\end{figure}

In the thermodynamic equation (Eq.~\ref{LatentHeat}), 
$Q_\theta$ is the rate of heating 
(expressed in ${\rm K}\,{\rm sec}^{-1}$) due to radiation. We adopt
a simple Newtonian relaxation scheme:
\begin{equation}
Q_\theta = -{{ \theta-\theta_{\rm ref}} \over {\tau_{\rm rad}}}
\label{newtonian}
\end{equation}
The equilibrium $\theta_{\rm ref}$ profiles, shown in Fig.~\ref{TP},
are based on pressure-temperature profiles following the radio-occultation and 
Galileo-probe results \citep{1981JGR....86.8721L, lindal-etal-1985,
lindal-etal-1987, lindal-etal-1992, 1998JGR...10322857S}. 
Each contains a deep neutrally stable troposphere, 
an isothermal stratosphere and a smooth transition layer 
between the two regions (Fig.~\ref{TP}).
The adopted relaxation timescales are 400 Earth days for Jupiter-type 
simulations and 200 Earth days for Neptune-type simulations.  These
timescales are shorter than expected radiative timescales in the deep
tropospheres of giant planets but allow us to perform simulations
in reasonable time while preserving the quasi-isentropic behavior of
the atmospheric motions over typical dynamical timescales of 1--10 days. 

Importantly, we chose to make $\theta_{\rm ref}$ independent of 
latitude for this study.  This contrasts with previous studies  
\citep[e.g.,][]{lian-showman-2008, Williams_2002, 
Williams_2003_2}, where the equilibrium temperature $\theta_{\rm ref}$ 
is a function of latitude.  Our choice here is motivated by
the fact that when $\theta_{\rm ref}$ depends
on latitude, the forcing imposes a zonally banded structure
on the flow, and it is thus unclear to what extent any zonal jet formation
result from such banded forcing rather than from the $\beta$ effect (where
$\beta$ is the gradient of Coriolis parameter with northward distance).
By making $\theta_{\rm ref}$ independent of latitude, we can ensure
that any banded flow structures result from $\beta$, not from
anisotropic forcing.  Moreover, when $\theta_{\rm ref}$ depends
on latitude, then not only the latent heating but the radiation
cause latitudinal temperature differences and thus injects
available potential energy (APE) into the flow.  The energy
source driving the flow would thus be ambiguous.  Here, we specifically
aim to test whether large-scale latent heating can drive Jovian-type
jets, and by maintaining $\theta_{\rm ref}$ constant with latitude,
we ensure that the only mechanism for generating lateral temperature
contrasts (hence APE) is large-scale latent heating.

The upper boundary in our simulations is zero pressure and impermeable.  
The lower boundary corresponds to an impermeable wall at a constant height; 
because the pressure can vary along this surface, it is implemented in pressure 
coordinates as a free surface through which no mass flow can occur 
(see \citet{campin2004} for details).  Both boundaries are free-slip in horizontal velocity 
The mean bottom pressure of simulated domain is a free parameter 
which varies from 100 to 500 bars depending on the simulation.
We adopt the ideal gas equation of state. The simulations include 
no explicit viscosity, but a fourth-order Shapiro filter \citep{Shapiro_1970} 
(analogous to eighth-order hyperviscosity) is added to maintain 
numerical stability. The time step is 100 sec.

Initially there are no winds in our simulations. The abundance of  
water vapor is set to be subsaturated ($\rm 95\%$ of saturation) in 
the region where $p<p_c$ and $q_{\rm deep}$ at $p>p_c$.    
In the initial condition, we introduce 5--9 random temperature 
perturbations at pressures less than $p_c$ to break the horizontal
symmetry and initiate motions. Each of the initial perturbations, which are 
positioned randomly within the simulated domain,  
has a warm center and affects the temperature radially within $\rm 5^\circ$.
The initial perturbations for our Jupiter, Saturn, and Uranus/Neptune
cases adopt $\Delta{\theta}=5$, 5, and 10 K, respectively, and
are confined to pressures less than 7, 10, and 10 bars, respectively.
We performed tests that varied the initial location and number of these perturbations,
which show that the qualitative final dynamical state, including the equatorial jet
direction, is not sensitive to the number of perturbations or their locations.
The main purpose of the perturbations
is to induce sufficient motion to generate supersaturation
in localized regions only at the very beginning of the simulations; 
once this occurs, the circulation becomes
self-generating.

Although the water abundances on Jupiter, Saturn, Uranus, and Neptune 
are unknown,
Galileo probe data indicate that Jupiter's C, N, S, Ar, Kr, and Xe 
abundances are all between 2--4 times solar.  
Spectroscopic information suggests that methane
is 7 times solar on Saturn \citep{flasar-etal-2005} and
30--40 times solar on Uranus and Neptune \citep{fegley-etal-1991, 
baines-etal-1995}. These values suggest that the water
abundance is modest at Jupiter, intermediate
at Saturn, and large at Uranus and Neptune.  Predicted 
condensation pressures are $\sim8\,$bars for Jupiter,
$\sim20\,$bars for Saturn, and 200--300 bars on Uranus 
and Neptune, depending on the water abundance 
\citep{flasar-etal-2005,fegley-etal-1991,baines-etal-1995}.

We explore a range of deep water-vapor abundances from 1--20 times
solar on Jupiter and Saturn and from 1--30 times solar on Uranus and Neptune. 
Combined with the prescribed temperature structure
(see Fig.~\ref{TP}), these abundances determine the range of pressures 
over which condensation will occur in any given simulation.  We then set $p_c$,
the pressure at the top of our deep water vapor source $Q_{\rm deep}$,
to be deeper than the base of the condensation region.
We use $p_c=7 \,\rm bars$ with 3 times solar  water abundance
and $p_c=10 \,\rm bars$ with 10 times solar water abundance for Jupiter 
simulations.  For Saturn, we used $p_c=17.3\,$bars for 5 times solar
and $19.2\,$bars for 10 times solar water abundance.
On Neptune, we use $p_c=120 \,\rm bars$ with 1 times solar water 
abundance, $p_c=220 \,\rm bars$ with 10 times solar water abundance and 
$p_c=330 \,\rm bars$ with 30 times solar water abundance. 
Here 1 times solar water abundance is 0.01 kilogram of water vapor per kilogram
of dry air.  Among all these simulations, Jupiter-type simulation with 
3 times solar water abundance and Neptune-type simulation with 30 times 
solar water abundance are the nominal cases. 

We run simulations using the cube-sphere grid.  Planetary parameters
are chosen based on Jupiter, 
Saturn, and Neptune, the latter of which represents the Uranus/Neptune
pair.  The parameters we implement for the nominal simulations are listed in
Table~\ref{parameter}, where
C128 stands for $\rm 128\times 128$ on each cubed-sphere face and 
C128 is equivalent to $\rm 512\times 256$ in 
longitude-latitude grid; C64 stands for $\rm 64\times 64$ on 
each cubed-sphere face and C64 is 
equivalent to $\rm 256\times 128$ in longitude-latitude grid; 
$N_L$ is number of layers in vertical direction.

\begin{table}
\begin{scriptsize}
\begin{tabular}  {|l|l|l|l|l|l|l|l|l|l|} 
\hline
\hline
Planet       & $a \rm{(km)}$  & $\Omega \rm{(s^{-1})}$  & $c_p \rm{(JK^{-1}kg^{-1})}$ &  ${\kappa}$ & $g \rm{(ms^{-2})}$ & ${R_q}$  & $Res$ & ${N_L}$ & $p_b \rm{(bars)}$\\ 
\hline
Jupiter       & 71492   & $\rm {1.7585\times 10^{-4}}$  & 13000   & 0.29    & 22.88  & -0.8778  & C128  &35 &100\\
Saturn       & 60268   & $\rm {1.6570\times 10^{-4}}$  & 13000   & 0.29   & 8.96    & -0.8778   & C128 & 35 & 100\\
Neptune   &  24746  & $\rm {1.0389\times 10^{-4}}$  & 13000   & 0.305   &11.7    &  -0.8778  & C64   &38 &500\\
\hline
\end{tabular}
\caption[Parameter table]
        {\label{parameter}
         Note: $a$ is radius of planet, $\Omega$ is rotation rate, 
$c_p$ is heat capacity, $\kappa={R \over {c_p}}$,  $g$ is gravity, 
$R_q={{1-\epsilon}\over \epsilon}$, where $\epsilon$ is the ratio of 
molecular mass between H$_2$O and H$_2$,  $Res$ is the horizontal 
resolution, $N_L$ is the number of layers, and 
$p_b$ is the mean bottom pressure of the simulated domain. }
\end{scriptsize}
\end{table}

\subsection{Diagnostics}

Before presenting our results, we describe the formalism we
use to diagnose our simulations following \citet{karoly-etal-1998}.  
For any quantity $A$, we can
define  $A=[A]+A^*$  where $[A]$ denotes the zonal mean and  $A^*$
denotes the deviation from the zonal mean.  Likewise, we can define
$A=\overline{A}+A'$, where $\overline {A}$ denotes the time average
and $A'$ denotes the deviation from the time average. Inserting these
definitions into the zonal momentum equation (Eq.~\ref{mom}) and
averaging in longitude and time, we obtain

\begin{eqnarray}
{\partial[\overline{u}]\over\partial t}=-{\partial\over\partial y}
([\overline{u'v'}] + [\overline{u}^*\overline{v}^*]) -
{\partial\over\partial p}([\overline{u'\omega'}] + [\overline{u}^*\overline{\omega}^*]) \nonumber \\
- [\overline{v}]{\partial [\overline{u}]\over\partial y} 
- [\overline \omega]{\partial [\overline{u}]\over\partial p} + f[\overline{v}]
\label{eulerian-mean-mom}
\end{eqnarray}

In Eq.~\ref{eulerian-mean-mom}, $[\overline{u'v'}]$ and $[\overline{u'\omega'}]$,
are the
latitudinal and vertical fluxes of eastward momentum, respectively, 
associated with traveling eddies;
$[\overline{u}^*\overline{v}^*]$ and $[\overline{u}^*\overline{\omega}^*]$,
are the latitudinal and vertical fluxes of eastward momentum, 
respectively, associated with stationary eddies. 
This equation states that latitudinal convergence
of horizontal eddy momentum flux, vertical convergence of vertical
eddy momentum flux, horizontal and vertical advection and Coriolis
acceleration drive the zonal winds. 

\section{Results}

\subsection{Basic flow regime}

\begin{figure}
 \centering
\includegraphics[width=5in]{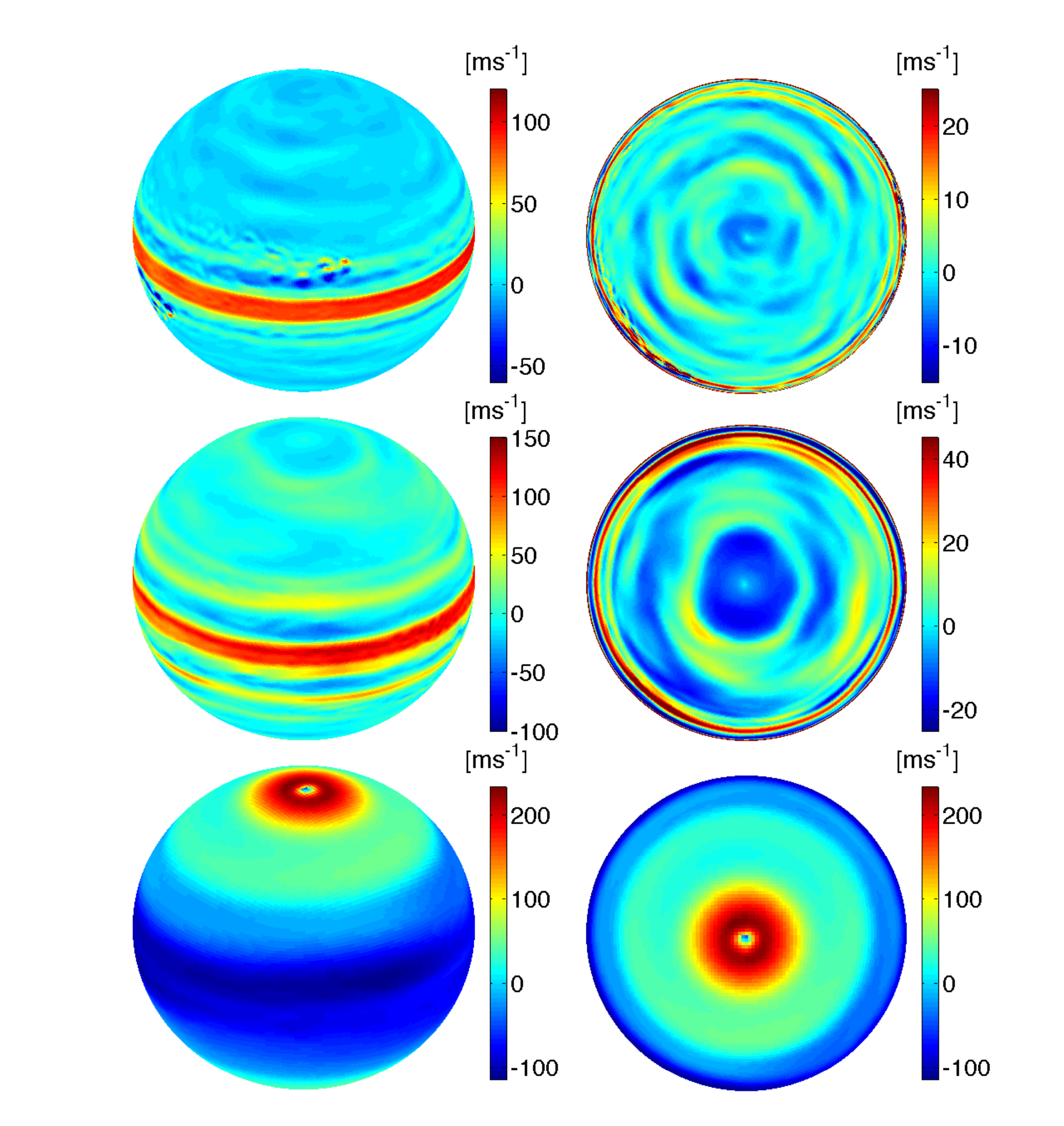}
\caption{Snapshots of zonal winds for Jupiter simulation (top row) 
with 3 times the solar water abundance at 0.9-bar level, Saturn simulation (center row) 
with 5 times the solar water abundance at 1-bar level and Uranus/Neptune 
simulation (bottom row) with 30 times the solar water abundance at 
the 0.8-bar level.  The simulation time for Jupiter and Uranus/Neptune  is 1200 Earth days. 
The simulation time for Saturn is 1600 Earth days. 
The left column gives an oblique view and the right column gives a view looking
down over the north pole. Our Jupiter and Saturn cases develop $\sim20$ jets with
equatorial superrotation while the Uranus/Neptune case develops 3
jets with equatorial subrotation.}
\label{jupiter_neptune_u_3d}
\end{figure}

Our Jupiter simulation with 3 times the solar water abundance
and Uranus/Neptune simulation with 30 times the solar water abundance
produce zonal winds  similar to those observed on Jupiter/Saturn and 
Neptune/Uranus.
The similarities, in general, are multiple banded zonal jets with
equatorial superrotation on Jupiter/Saturn and high-latitude  eastward
jets with broad equatorial subrotation on Neptune/Uranus.  These are shown in
Fig.~\ref{jupiter_neptune_u_3d}.   
The initial perturbations generate motion, which triggers condensation.
Once the circulation is initiated, it becomes
self-sustaining: horizontal temperature gradients induced by
large-scale latent heating drive a circulation that continues to dredge up
water vapor to the condensation region, allowing latent heating
and maintaining the temperature differences.  The eddies produced
in this way interact with planetary rotation to generate large-scale zonal 
flows. The resulting zonal flow at 
the 1-bar level contains about 20 zonal jets for Jupiter/Saturn-type
simulations and $\sim3$ zonal jets for Uranus/Neptune-type
simulations (Fig.~\ref{jupiter_neptune_u_3d}). 

\begin{figure}
 \centering
\includegraphics[width=5in]{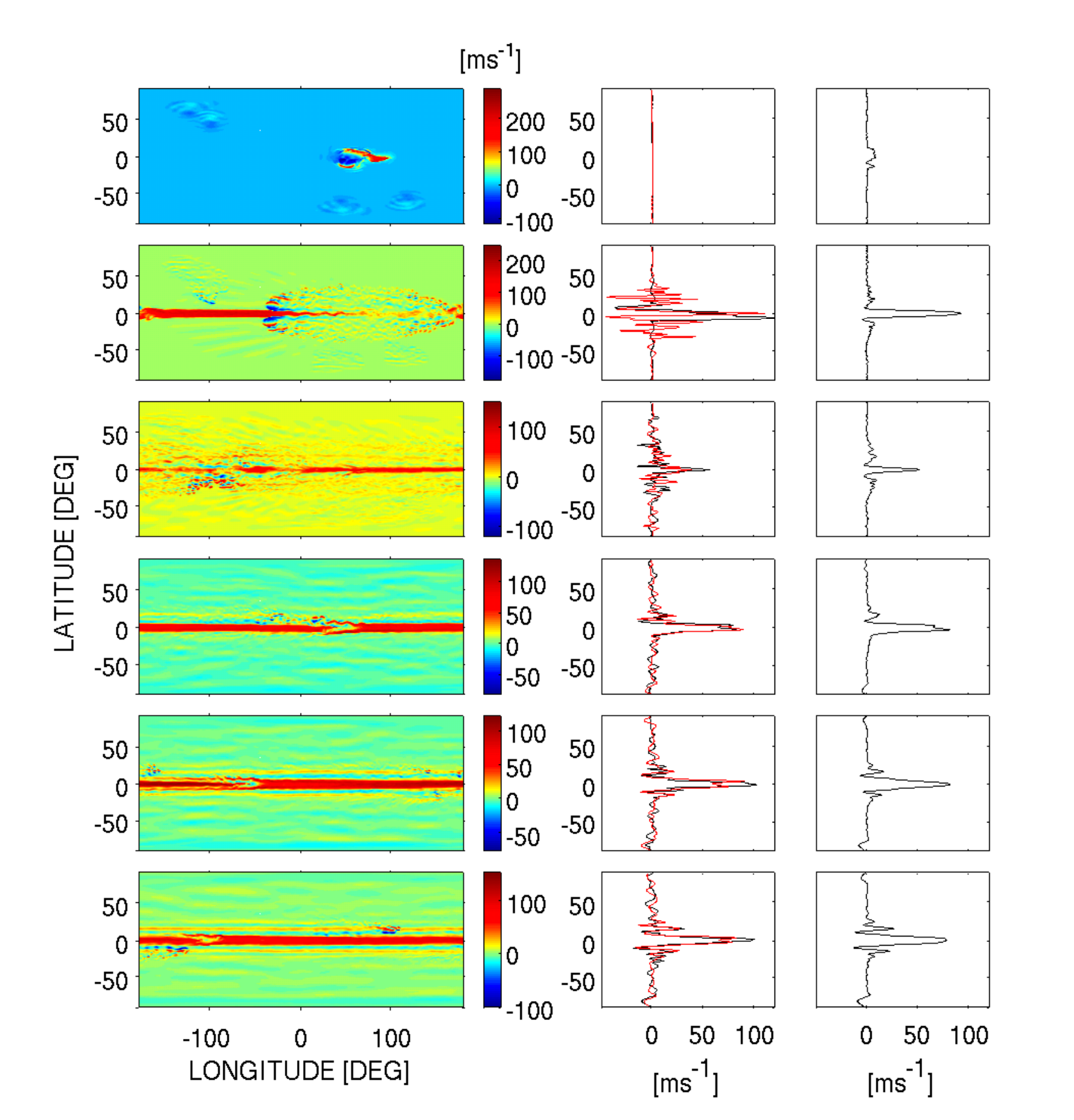}
\caption{Snapshots of zonal winds at 0.9 bars for our Jupiter 
simulation with 3 times the solar water abundance. 
The four rows, from top down, give the state at 3, 55, 116, 578, 1157 and 2315 
Earth days.  The left column shows the zonal wind over the full globe
in a rectangular projection.  The middle column gives the zonal wind 
profile at longitude $\rm -176.8^\circ$ (black line) and 
$\rm 0^\circ$ (red line). The right column is zonal-mean zonal wind.}
\label{jupiter_3s_evol}
\end{figure}

First we examine the Jupiter simulations (Fig.~\ref{jupiter_3s_evol}).  
By 55 Earth days, the winds show significant zonality, and after $\sim1100\,$days 
the jet pattern becomes relatively stable 
with $\sim20$ jets.  The equatorial jet builds up rapidly,
reaching zonal-mean zonal wind speeds of  $80\,{\rm m}\,{\rm sec}^{-1}$
and local speeds exceeding $100\,{\rm m}\,{\rm sec}^{-1}$.
Initially, the jet spans only a range of longitudes (e.g.,
Fig.~\ref{jupiter_3s_evol}, second panel).  Low-latitude eddies 
triggered by localized latent heating continuously
pump energy into the equatorial flow, which eventually makes the
equatorial superrotation encircle the whole globe by 116 Earth days.
The longitudinal variation of this equatorial superrotation becomes 
small after $\sim1000$ Earth days.  The jet 
spans latitudes $\rm 10^\circ$ south to $\rm 10^\circ$ north with an average
wind speed of $\rm 80ms^{-1}$.

\begin{figure}
 \centering
\includegraphics[width=6in]{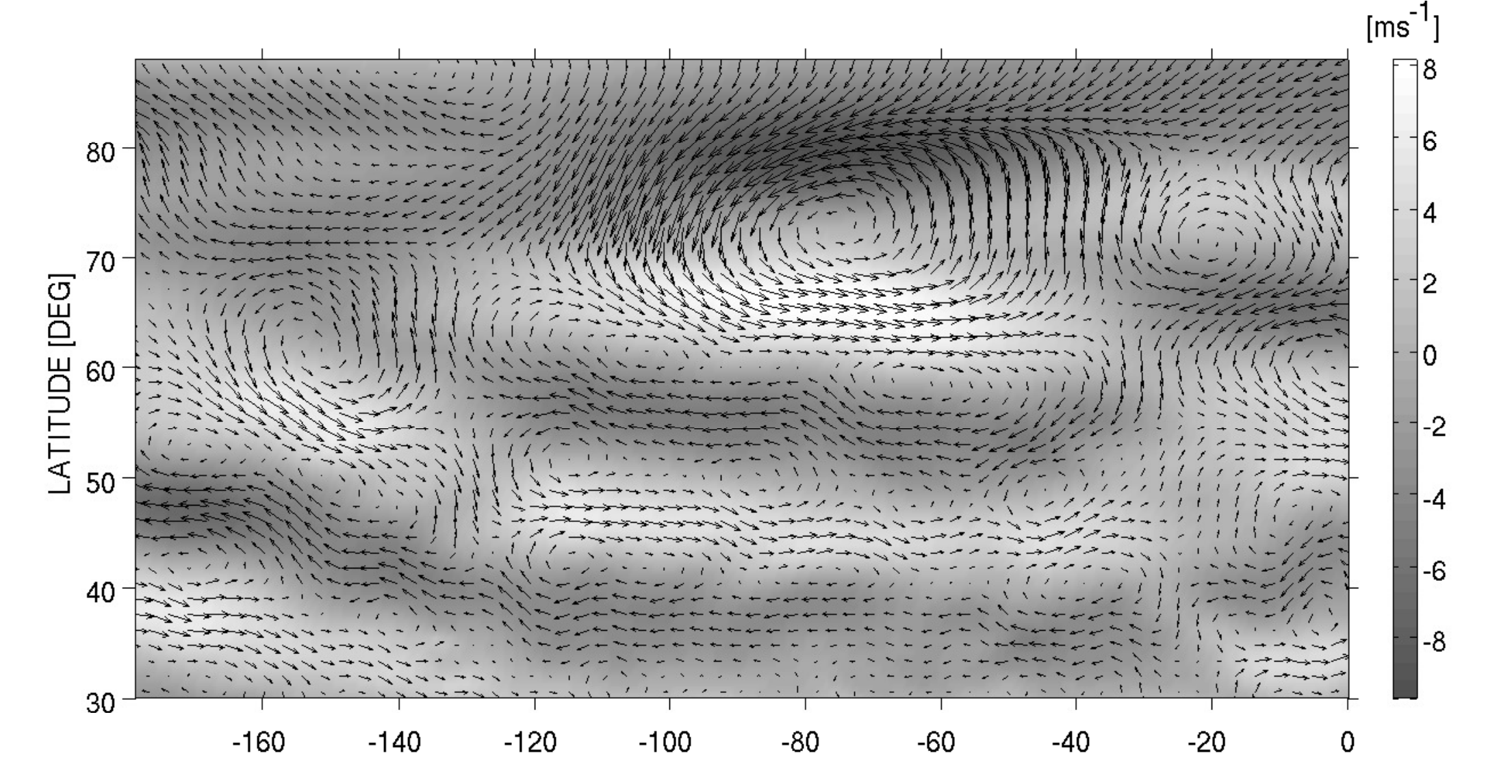}
\caption{A zoom-in showing the wind structure in a portion of
the domain for our Jupiter case with 3 times solar water.
Gray scale gives zonal winds and arrows show full wind velocity
at 0.9 bars.  This figure is at 2315 Earth days. The wind field shown here 
is in longitude-latitude section from $\rm -180^\circ$ to $0^\circ$ 
longitude and from $\rm 30^\circ$ N to $90^\circ$N latitude.}
\label{jupiter_uv_vector}
\end{figure}

In our Jupiter simulations, numerous alternating east-west jets
also develop at higher latitudes with speeds of 
$\sim5$--$10\,{\rm m}\,{\rm sec}^{-1}$.  
These high-latitude zonal jets extend almost to the pole 
(Fig.~\ref{jupiter_neptune_u_3d}).
Interestingly, however, the high-latitude jets are 
not purely zonal but develop meanders with latitudinal positions that vary 
in longitude, as can be seen in Figs.~\ref{jupiter_neptune_u_3d} and
\ref{jupiter_uv_vector}.
This meandering presumably occurs because the $\beta$ effect (which
is necessary for jet formation) weakens
at high latitudes.  As a result of these meanders, a zonal average
smoothes through these jets, so the zonal-mean
zonal wind profile shows minimal structure poleward of $\sim30^{\circ}$
latitude (Fig.~\ref{jupiter_3s_evol}, rightmost panels); nevertheless,
the high-latitude jet structure remains evident in profiles without
zonal averaging (Figs.~\ref{jupiter_neptune_u_3d} and \ref{jupiter_3s_evol}). 
{\t Interestingly, in the Saturn case, an eastward jet at $\sim 70^{\circ}$ latitude
develops meanders crudely resembling a polygon when viewed from over the pole 
(Fig.~\ref{jupiter_neptune_u_3d}).  This structure may have relevance to explaining
Saturn's polar hexagon \citep{godfrey-1988, baines-etal-2009}.}

\begin{figure}
 \centering
\includegraphics[width=5in]{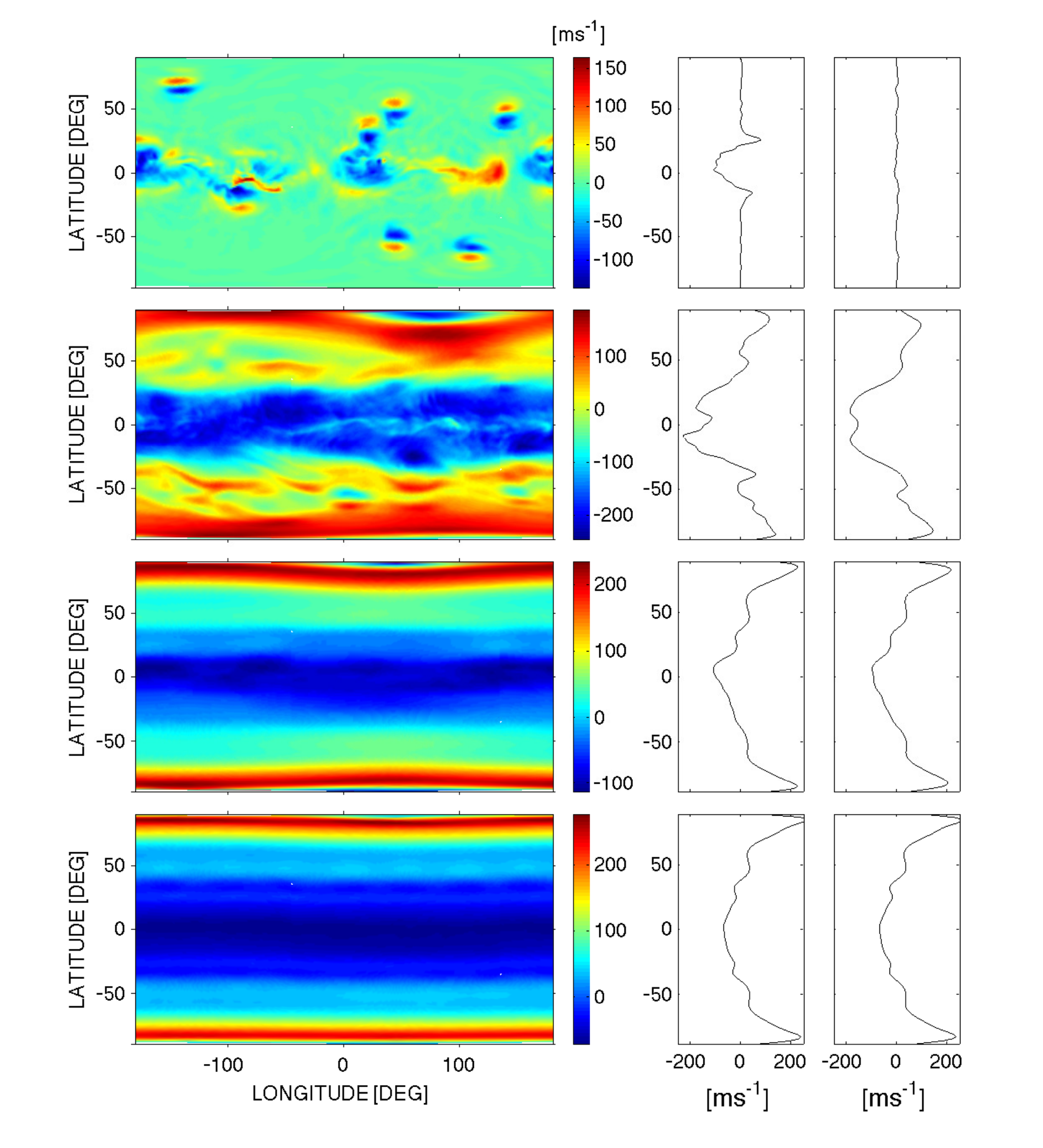}
\caption{Snapshots of zonal winds at 0.8 bars for our Uranus/Neptune 
simulation with 30 times the solar water abundance.  The four rows, 
from top down, give the state at 3, 55, 1157 and 2315 Earth days.  
The left column shows the zonal wind over the full globe in a 
rectangular projection.  The middle column gives the zonal wind profile
at longitude $-176.8^\circ$. The right column gives the zonal-mean zonal wind.}
\label{neptune_30s_evol}
\end{figure}

Our Uranus/Neptune simulation with 30 times the solar water abundance, 
however, behaves quite differently than our Jupiter and Saturn cases
(Fig.~\ref{neptune_30s_evol}). 
By $\sim1000$ days, the profile stabilizes with three jets:
a broad westward equatorial flow extending from latitudes $40^\circ$ 
south to $40^\circ$ north and reaching speeds of almost $-100
\,{\rm m}\,{\rm sec}^{-1}$, and two high-latitude eastward jets
reaching peak speeds of almost $250\,{\rm m}\,{\rm sec}^{-1}$
at latitudes of 70--$80^{\circ}$ north and south.
Eddy activity, though still vigorous, is hardly visible in comparison 
with the zonal flow after several hundred Earth days.

As a control experiment, we also performed a Neptune simulation where latent heating
and condensation were turned off (i.e., $\delta=0$ in Eqs.~\ref{LatentHeat}--\ref{Condensation}
regardless of the relative humidity).  Because $\theta_{\rm ref}$ is independent
of latitude, radiation {\it removes} rather than adds available potential
energy, and thus the only source of available potential energy in this 
simulation is provided by the thermal perturbations in the initial condition.  Consistent
with this expectation, this simulation develops peak winds of only $\sim20\rm\,m\,sec^{-1}$,
an order-of-magnitude weaker than those our nominal simulation.  This comparison demonstrates 
the crucial role that latent heating plays in generating jets in our nominal simulations.

\begin{figure}
 \centering
\includegraphics[width=5in]{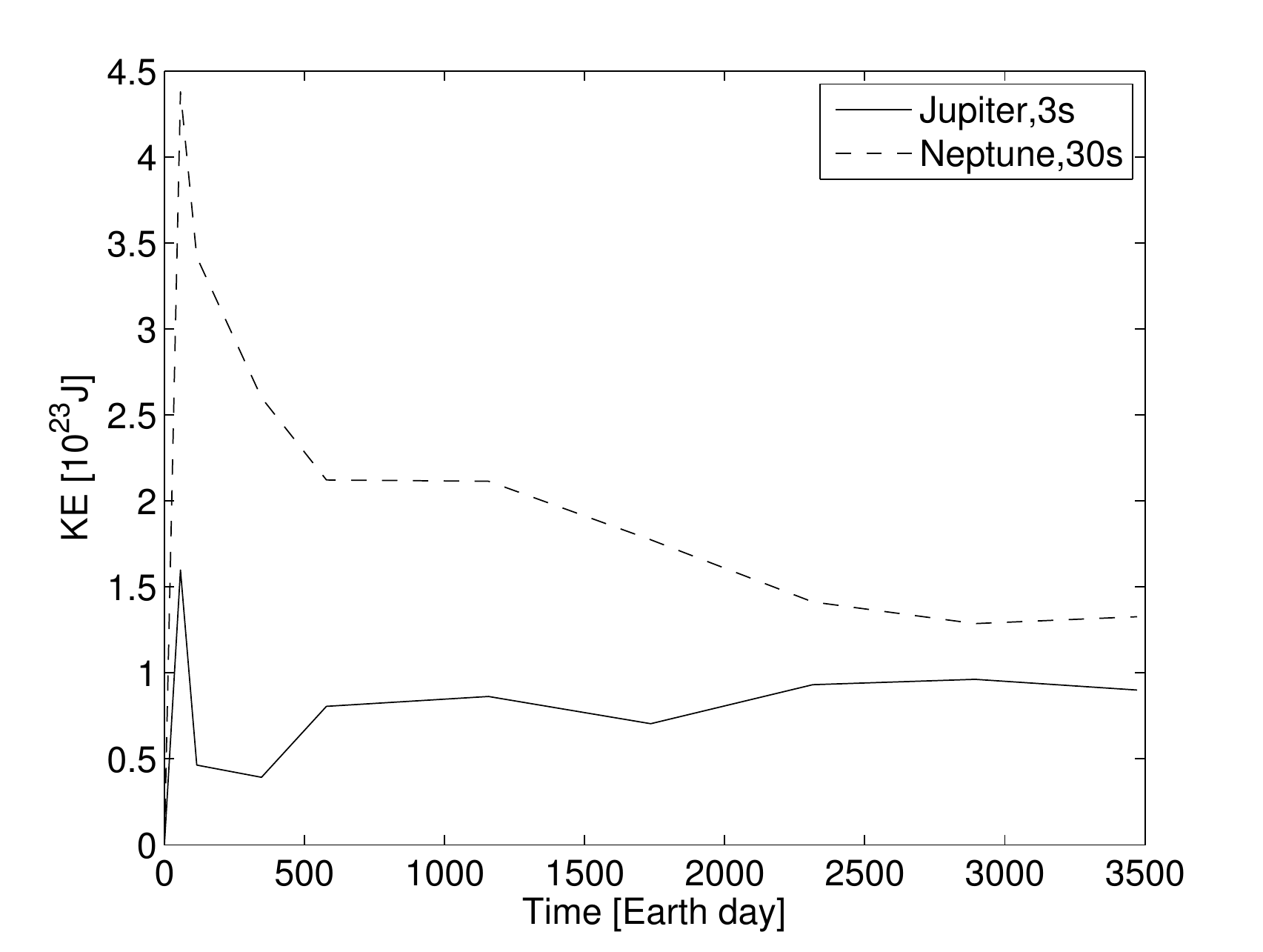}
\caption{Total kinetic energy in region from 1 bar and above as a function of 
time. The solid line shows our nominal Jupiter simulation with 3 times the solar 
water abundance and the dashed line shows nominal Neptune simulation with 30 times the 
solar water abundance. }
\label{JN_KE}
\end{figure}

Figure.~\ref{JN_KE} shows the time evolution of kinetic energy in 
our Jupiter and Uranus/Neptune simulations. The kinetic energy is vertically 
and horizontally integrated 
in region from 1 bar and above. Both Jupiter and Uranus/Neptune simulations 
show that kinetic energy quickly spikes up in first 100 Earth days and 
gradually drops afterwards. After 2500 Earth days, the variation of 
kinetic energy becomes small, indicating the simulations get close to 
a steady state from top down to 1 bar. This development of kinetic energy 
is very similar to that of \citet{lian-showman-2008}. Nevertheless, the 
barotropic winds continue to spin up at deep levels near the bottom of
the model.
 
Our simulations provide a possible explanation for the 
equatorial superrotation on Jupiter/Saturn yet the equatorial 
subrotation on Uranus/Neptune as well as the approximate number of jets
observed on all four planets. 
 We emphasize that our simulated jet 
profiles --- including the equatorial jet direction ---
are fully self-generating and emerge spontaneously, without the 
application of {\it ad hoc} forcing schemes.  The only physical
differences between the two simulations in Fig.~\ref{jupiter_neptune_u_3d} 
is the values of the planetary parameters (radius, rotation
rate, gravity) and deep water abundance $q_{\rm deep}$;
the forcing schemes are otherwise identical for the two cases.
In contrast, previous shallow-atmosphere studies either
produced superrotation only with artificially imposed forcing near the equator 
\citep{Williams_2003_1, 2005P&SS...53..508Y, lian-showman-2008} or
produce superrotation more naturally but make no prediction for 
Jupiter/Saturn versus Uranus/Neptune \citep{scott-polvani-2008, schneider-liu-2009}.
Ours is the first study to naturally produce superrotation in a Jupiter regime
yet subrotation in a Uranus/Neptune regime within the context
of a single model.

We emphasize that the jet widths that emerge in our simulations 
are self-selecting; neither the scales of zonal jets nor the
scales of baroclinic eddies are controlled by the initial perturbations. 
Our Jupiter-type simulation with 3 times solar water abundance has 
jet widths ranging from several thousand kilometers at mid-to-high 
latitudes to about $\rm 15,000$ kilometers 
at the equator.  
Our Uranus/Neptune simulation with 30 times solar water abundance 
has jet widths of around 25,000  km. These 
jet widths are similar (within a factor of $\sim 2$) 
to the Rhines scale $\pi (2U/\beta)^{1/2}$, 
where $U$ is the characteristic jet speed.

\begin{table}
\begin{scriptsize}
\begin{tabular}  {|l|l|l|l|l|l|l|} 
\hline
\hline
Planet       & $\frac{a}{a^\circ}$  & $\frac{\Omega}{\Omega^\circ}$ & $\frac{q_{deep}}{q_{solar}}$ & $Res$ & ${N_L}$ &$p_b \rm{(bars)}$ \\ 
\hline
Saturn       & 0.5 & 0.5, 1, 2  &  5  & C128  &35 &100\\
Saturn       &   1  & 0.5, 1, 2  &  5  & C128  &35 &100\\
Saturn       &   2  & 0.5, 1, 2  &  5  & C128  &35 &100\\
Saturn       & 0.5 & 0.5, 1, 2  & 20 & C128  &35 &100\\
Saturn       &   1  & 0.5, 1, 2  & 20 & C128  &35 &100\\
Saturn       &   1  &        1      &  1  & C128  &35 &100\\
Saturn       &   1  &        1      &  3  & C128  &35 &100\\
Saturn       &   1  &        1      & 10 & C128  &35 &100\\
Saturn       &   1  &        1      & 20 & C128  &35 &100\\
Neptune    &   1  &        1      &  1  & C64    &38 &500\\
Neptune    &   1  &        1      &  3  & C64    &38 &500\\
Neptune    &   1  &        1      & 10 & C64    &38 &500\\
\hline
\end{tabular}
\caption[Parameter table for selected test cases]
        {\label{parameter_variation}
         Note: $\frac{a}{a^\circ}$ and $\frac{\Omega}{\Omega^\circ}$ are ratios between planet radius 
          and rotation rate in test cases and those in nominal cases listed in table~\ref{parameter} respectively. 
          $\frac{q_{deep}}{q_{solar}}$ is the ratio between deep water abundance in test cases and solar water abundace. 
          We maintain $c_p$, $\kappa$,  $g$ and $R_q$ to be the same as those in the nominal cases. 
}
\end{scriptsize}
\end{table}

\begin{figure}
 \centering
\includegraphics[width=5in]{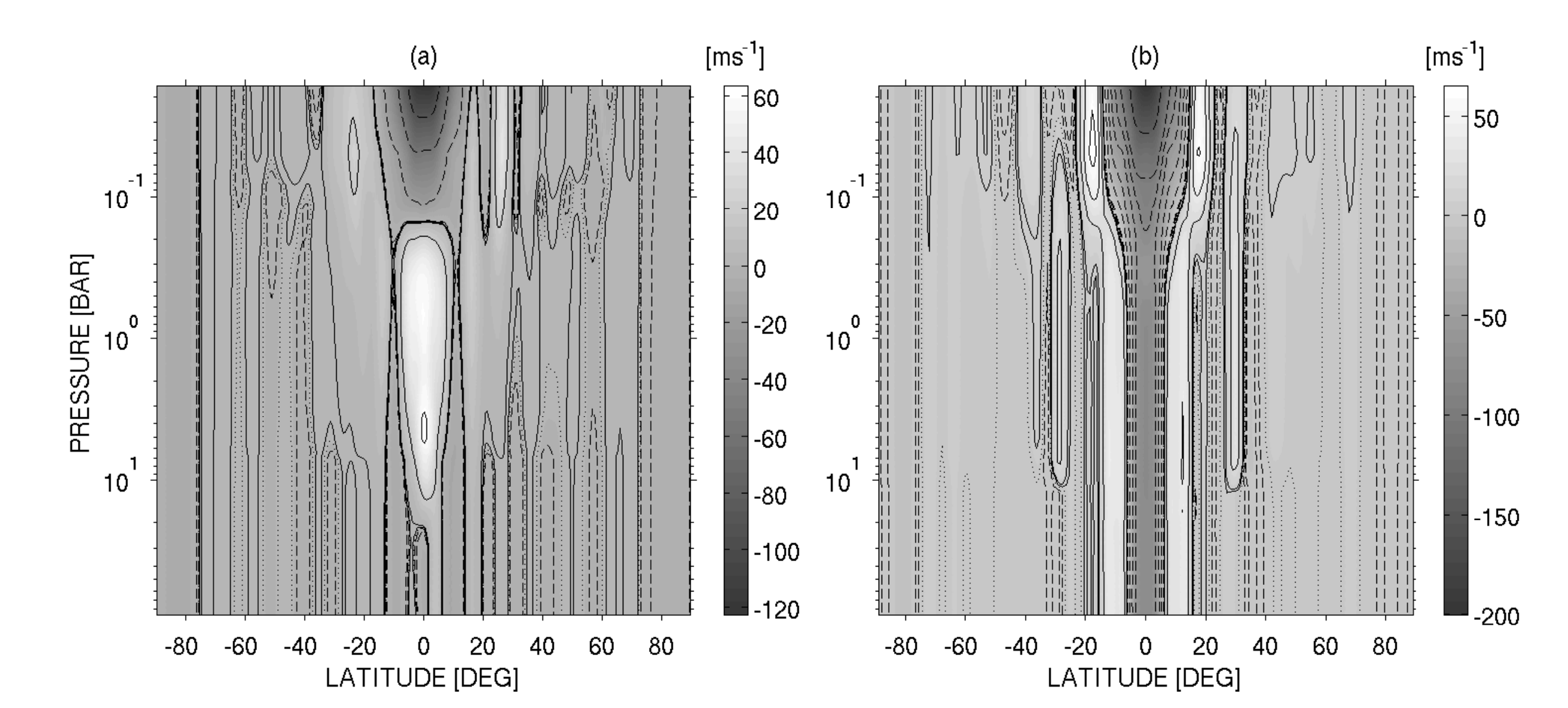}
\caption{Zonal-mean zonal winds for Jupiter simulations with 
3 times solar water abundance (left) and 10 times solar water
abundance (right) at 1157 Earth days.  The structures were time averaged
for 231 Earth days. The solid and dashed lines depict eastward and
westward winds, respectively, and the dotted lines represent zero speed. 
These lines only indicate the direction of the zonal winds.  Note
the development of equatorial superrotation in the 3-times-solar-water
case and equatorial subrotation in the 10-times-solar-water case. }
\label{jupiter_3s_10s_u}
\end{figure}

\begin{figure}
 \centering
\includegraphics[width=5in]{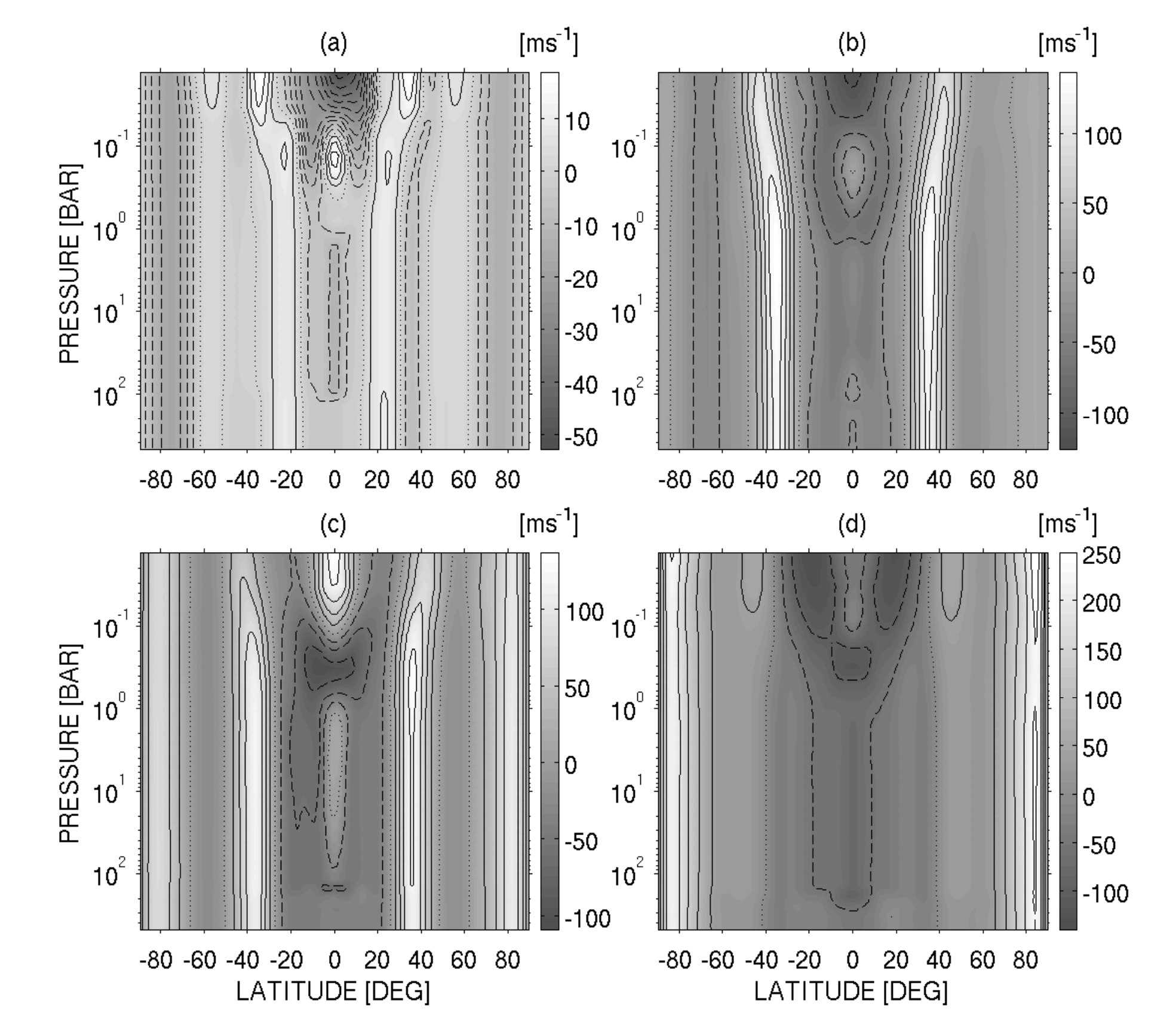}
\caption{Zonal-mean zonal winds for Uranus/Neptune simulations 
with various abundances of water vapor at 2200 Earth days. The 
structures were time averaged for 231 Earth days. The solid and dashed
lines depict eastward and westward winds, respectively, and the 
dotted lines represent zero speed.
(a) 1 times the solar water abundance , line spacing is $\rm 5ms^{-1}$, (b) 3 times the solar water abundance, 
line spacing is $\rm 30ms^{-1}$, 
(c) 10 times the solar water abundance, line spacing is $\rm 30ms^{-1}$ and (d) 30 times the solar water abundance, 
line spacing is $\rm 50ms^{-1}$.  Note that equatorial superrotation 
develops at low water abundances, but equatorial
subrotation develops at 30 times solar
water abundance.}
\label{Neptune_U_cs}
\end{figure}

To investigate the influence of water abundance on the circulation,
and to shed light on what causes the differences between our
Jupiter/Saturn and Uranus/Neptune cases  (Fig.~\ref{jupiter_neptune_u_3d}),
we ran a series of simulations exploring a range of deep water-vapor
abundances (Table~\ref{parameter_variation}).  This is carried out simply by varying the value of
the deep water abundance, $q_{\rm deep}$, adopted in our water-vapor
source term $Q_{\rm deep}$.  Figure~\ref{jupiter_3s_10s_u} 
shows the results for Jupiter cases with 3 and 10 times solar water
while Fig.~\ref{Neptune_U_cs} shows the results for Uranus/Neptune cases
with 1, 3, 10, and 30 times solar water.   Interestingly, we find
in both cases that equatorial superrotation
preferably forms at low water-vapor abundance while subrotation forms
at high water-vapor abundance.  For Jupiter, 3-times-solar water
yields superrotation while 10-times-solar water produces subrotation.
For Uranus/Neptune, solar water (panel {\it a}) produces a narrow
superrotating jet centered at pressures of $\sim100$--400 mbar;  
at 3-times-solar water (panel {\it b}), a local maximum in zonal
wind still exists at that location, but its peak speeds are slightly
subrotating.  The ten-times-solar case (panel {\it c}) bucks the
trend, developing superrotation in the lower stratosphere 
(pressures $<100\,$mbar).  By 30 times solar, however, the structure
becomes more barotropic and the equatorial jet direction is subrotating
throughout. Nevertheless, the superrotation in our low-water-abundance
Uranus/Neptune cases is weaker than that in our Jupiter/Saturn cases,
which suggests that water abundance is not the only factor that controls 
the existence of superrotation in our simulations.  We return to this issue 
in Section \ref{parameter-variation}.

We also performed Saturn simulations with not only 5-times-solar water 
abundance (as seen in Fig.~\ref{jupiter_neptune_u_3d}) but 10 and 20 times 
solar as well. All these cases developed multiple banded zonal jets at 
both low and high latitudes.  The 5-times-solar 
Saturn case developed equatorial superrotation, with zonal-mean speeds reaching
$\sim150\rm\,m\,s^{-1}$ eastward, while the 10-times-solar and 20-times-solar
cases developed equatorial subrotation with speeds reaching 
$-100\rm\,m\,s^{-1}$ or more westward near 1 bar.  While our ability to produce 
superrotation in a Saturn case with 5-times-solar water is encouraging, Saturn's
water abundance is unknown and could easily be as high as 10 times solar
\citep[e.g.,][]{mousis-etal-2009}; moreover, even our superrotating case 
produced a superrotating jet much weaker and narrower than 
the observed jet (which extends from $30^{\circ}$N to
$30^{\circ}$S latitude and reaches peak speeds of $\sim400\rm\,m\,s^{-1}$).  
This disagreement could mean that Saturn's 
equatorial jet does not fit into the framework discussed here, and 
that the observed jet results instead from a different mechanism. 
On the other hand, processes excluded here, including moist convection,
evaporating precipitation, and realistic radiative transfer, will all
affect the tropospheric static stability and thus could influence the
eddy/mean-flow interactions that pump the equatorial jet.
Definitely assessing whether Saturn's equatorial
jet can result from cloud-layer processes will thus require next-generation
models that include these improvements.

Another trend that occurs in our simulations is that the 
mean jet speeds and jet widths increase as the deep water-vapor
abundance is increased.  As a result, simulations with less
water tend to have more jets than simulations with greater water.
This trend is most evident in our Uranus/Neptune cases in
Fig.~\ref{Neptune_U_cs}: the peak jet speed increases from
$\sim50\,{\rm m}\,{\rm sec}^{-1}$ to $250\,{\rm m}\,{\rm sec}^{-1}$
as water is increased from solar to 30 times solar, while
the number of jets drops from seven to three over this same sequence.
However, the case with 10-times-solar-water  
does not fit the trend well. It has 7 zonal jets, the same as
our solar case, and its wind speeds are similar to that of our
3-times-solar case.

\begin{figure}
 \centering
\includegraphics[width=4in]{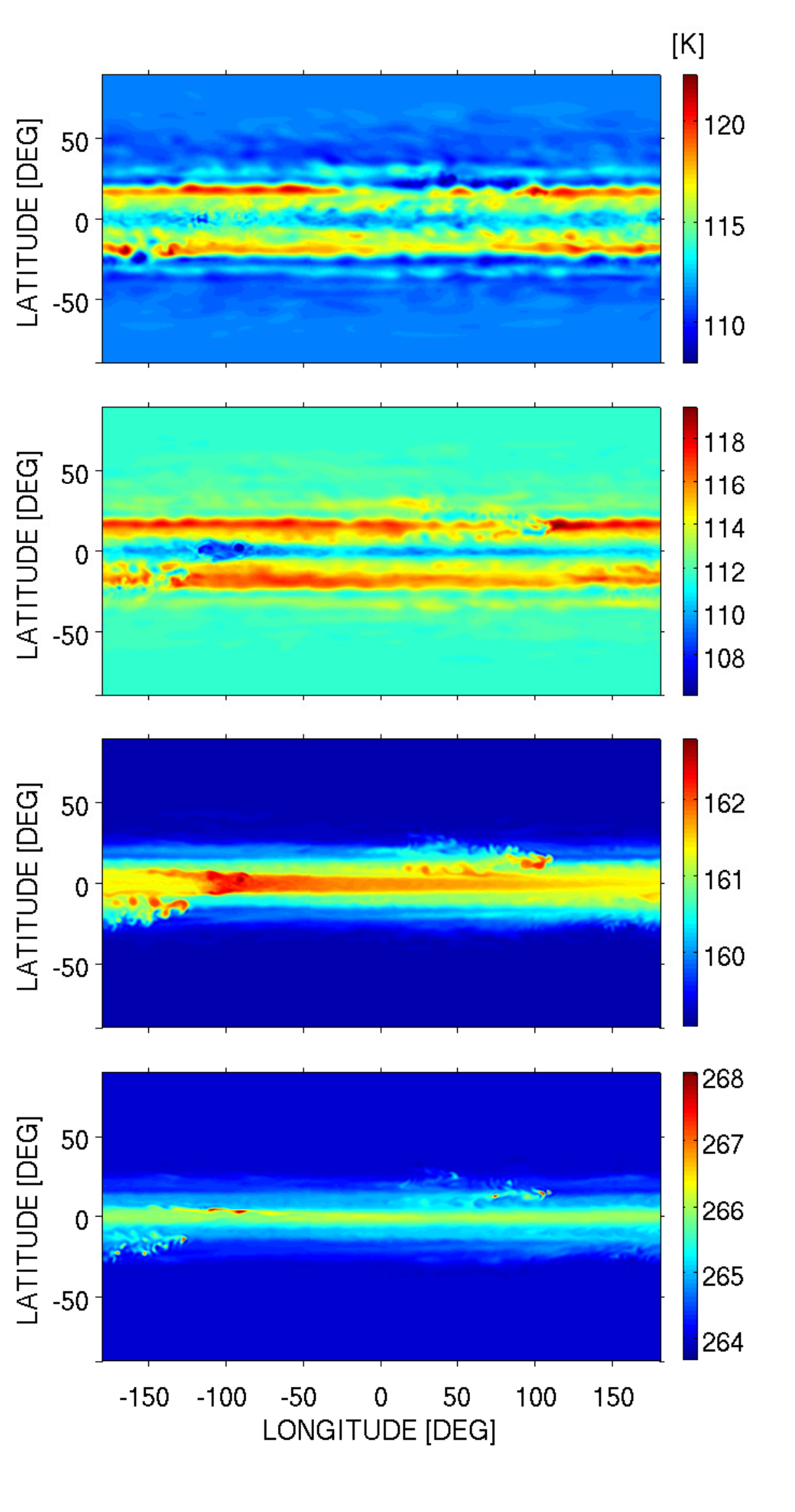}
\caption{Global temperature maps for Jupiter simulation with 3 times 
the solar water abundance at 2315 Earth days. 
The pressure levels are, from top down, 0.1-bar level, 0.2-bar level, 0.9-bar level and 5-bar level.}
\label{jupiter_T}
\end{figure}

Figure~\ref{jupiter_T} shows the temperature structure in our
nominal Jupiter simulation at pressures of 0.1, 0.2, 0.9, and 5 bars
from top to bottom, respectively.  The top two panels are in the
lower stratosphere and upper troposphere; the third panel is near
the top of the region with strong eddy accelerations, and the bottom
panel is just above the water condensation level.  Interestingly,
in the deep regions where latent heating occurs (bottom two panels), 
the latitudinal temperature contrasts primarily occur within
$\sim20^{\circ}$ of the equator.  In the simulations of 
\citet{Williams_2003_1} and \citet{lian-showman-2008},
equatorial superrotation developed only in the presence of 
large latitudinal temperature contrasts near the equator; this lead to a 
barotropic instability that pumped energy and momentum into the
superrotating jet.  In their simulations, these near-equatorial 
temperature gradients resulted from an {\it ad hoc}
Newtonian heating profile.  Here, however, these sharp near-equatorial
temperature gradients develop naturally from the interaction of the
moist convection with the large-scale flow.  Despite this difference,
the similarity of the resulting near-equatorial temperature profiles 
suggest that the superrotating jet-pumping mechanism identified by 
\citet{Williams_2003_1} and \citet{lian-showman-2008} could be
relevant here.  
In the upper troposphere and lower stratosphere, our Jupiter simulation
develops a banded temperature pattern, with latitudinal temperature
differences of $\sim10\,$K, that bears some similarities to that
observed on Jupiter and Saturn.  Interestingly, the simulated equatorial
zone is cold in the upper troposphere, consistent with observations
of Jupiter and Saturn; this temperature structure is associated with
the decay of the equatorial jet with altitude (see 
Fig.~\ref{jupiter_3s_10s_u}) via thermal-wind balance.  
Several regions of localized latent heating and eddy generation
are visible in both hemispheres; Section~\ref{storms} 
presents a detailed discussion of these features.

\begin{figure}
 \centering
\includegraphics[width=5in]{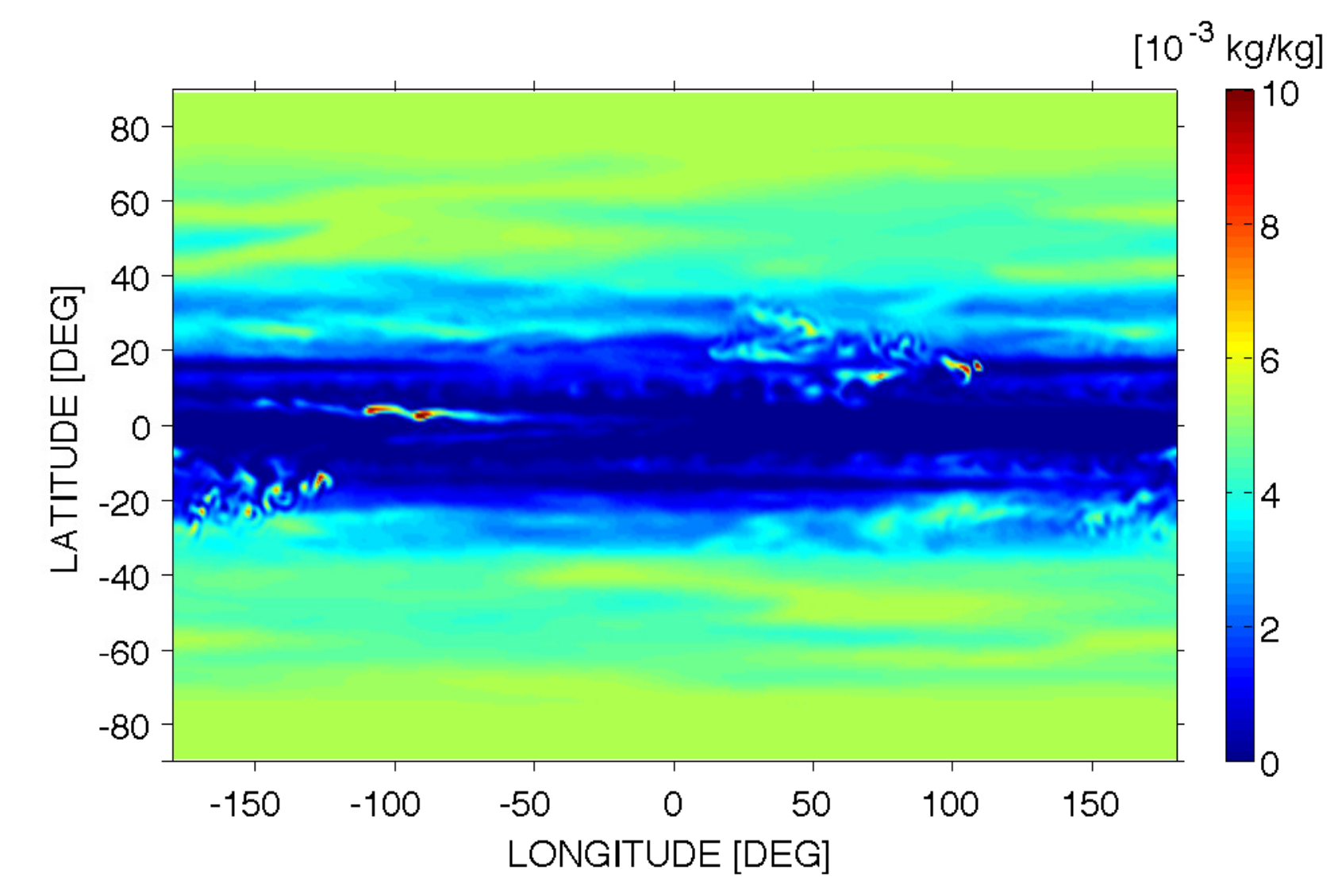}
\caption{Global map of water-vapor mixing ratio 
for Jupiter simulation with 3 times 
the solar water abundance at 2315 Earth days. Shown at the 
5-bar level.}
\label{jupiter_S}
\end{figure}

Figure~\ref{jupiter_S} depicts the distribution of water vapor at 
the 5-bar level for our nominal Jupiter simulation at the same time
as the temperature plots shown in the previous figure.  A zonally banded structure
is evident,
with low water vapor near the equator and intermediate values at mid-to-high
latitudes.  Localized regions of latent heating manifest as
regions of high water vapor abundance near the equator and at $\sim20^{\circ}$
latitude in both hemispheres (orange/red regions in the figure).  While the connection
between temperature and water vapor is visually evident (compare Figs.~\ref{jupiter_S} 
and \ref{jupiter_T}, bottom panel), scatter plots of temperature versus water-vapor 
mixing ratio at a given isobar show a great deal of scatter, indicating that no 
simple relationship connects the two quantities.  Diagnostics of the mechanisms that
determine the water and temperature distributions will be presented in a future paper.

\subsection{What controls the equatorial flow}
\label{parameter-variation}
Among all our simulation results, the most interesting feature is that
the direction and strength of equatorial flow vary with 
deep water abundance; eastward equatorial flow forms at low deep water 
abundance and westward equatorial flow forms at high deep water abundance. 
What causes this trend? Moreover, the planets in our simulations have 
different radii and rotation rates. Can these affect the formation of 
equatorial flow? Here we address these questions. 

\begin{figure}
\centering
\includegraphics[width=4in]{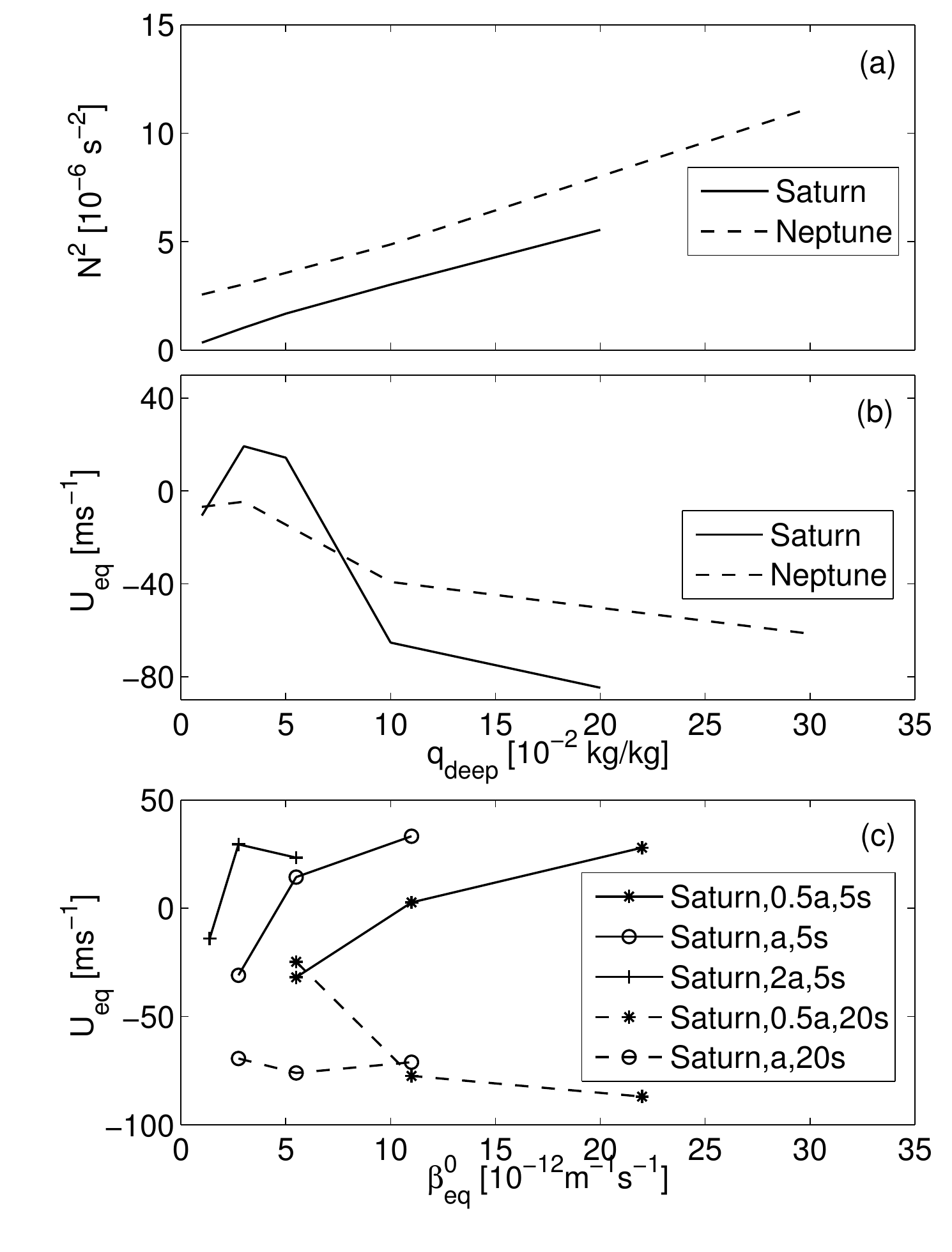}
\caption{ (a) Relation between $N^2$, where $N$ is Brunt Vaisala 
frequency in region deeper than 1 bar, and $q_{\rm deep}$
for Saturn and Neptune cases.  Greater water abundance leads to
greater static stability in the troposphere.  
(b)  Relation between height-averaged
equatorial wind and deep water abundance for Saturn and Neptune
cases.  Greater water abundance generally leads to stronger
westward equatorial wind. (c) Relation between
height-averaged equatorial wind and equatorial value of $\beta$,
the gradient of Coriolis parameter (equal to $2\Omega/a$ at the
equator) in Saturn-type cases varying the planetary rotation rate,
radius, and deep water abundance. The relationship is complex (see
text).  Solid lines show 5 times solar
and dashed lines show 20 times solar $q_{\rm deep}$.  Each curve
varies the rotation rate at a given $q_{\rm deep}$ and radius 
(between half and
double Saturn's actual radius, as marked in the legend), and
different curves use different radii and/or $q_{\rm deep}$.
All results are shown at 1157 Earth days.  
Table~\ref{parameter_variation} lists all the cases shown here.}
\label{parameter_sweep}
\end{figure}

We performed additional test cases by varying the deep-water-vapor abundance, 
planetary radius, and planetary rotation rate (Table~\ref{parameter_variation}). 
By comparing the equatorial zonal wind, Brunt Vaisala frequency, and $\beta$,
we seek to reveal the major factor that affects the trend. In the following 
discussion, the zonal wind is vertically averaged from the bottom to top of 
the simulated domain, while the Brunt-Vaisala frequency is vertically averaged 
from the bottom to 1 bar to exclude the large static stability in the stratosphere 
and thus better demonstrate the effects of latent heating in the troposphere. 
First, we vary the deep water abundance for Saturn and Neptune 
simulations using the nominal planetary radii and rotation rates; the
results are shown in the top two panels of Fig.~\ref{parameter_sweep}. 
Figure.~\ref{parameter_sweep}(a) clearly shows that when the deep 
water abundance increases, the tropospheric Brunt-Vaisala frequency increases. 
Figure~\ref{parameter_sweep}(b) confirms that larger deep water abundance
(hence tropospheric static stability) makes the equatorial flow more
westward.  This correlation applies to both the Saturn and Neptune simulations 
(solid and dashed lines and dashed lines in Fig.~\ref{parameter_sweep}, respectively) 
except the case of Saturn with 1-times-solar water abundance, in which the equatorial 
flow is very weak due to the low deep water abundance. 

Although a clear correlation exists between greater deep water abundance
and faster westward equatorial flow (Fig.~\ref{parameter_sweep}b), the Neptune
simulations exhibit a different dependence than the Saturn simulations.
This suggests that other factors, such as planetary radius or rotation rate, 
play a role in controlling the equatorial jet speed.   In an attempt to untangle these effects,
we ran Saturn sensitivity studies that varied the planetary radius or rotation 
rate from nominal Saturn values but kept all other parameters fixed.  Cases with 
deep water abundances of 5 and 20 times solar were explored.  Figure~\ref{parameter_sweep}(c) 
depicts the equatorial wind speed versus the equatorial value of $\beta$, the gradient of
the Coriolis parameter (just $2\Omega/a$ at the equator).
Solid lines denote the 5-times-solar cases while the dashed lines show the 20-times-solar
cases.  Different symbols denote different planetary radii and/or deep-water abundances;
lines connect sequences of simulations performed at a given planetary radius and
deep-water abundance but with differing rotation rate.
At 5-times-solar water abundance, increasing the rotation rate (at constant planetary radius)
makes the equatorial flow more eastward.  However the situation reverses at 20-times-solar,
where increasing the rotation rate either has minimal effect on the equatorial jet
(for Saturn's radius) or makes the jet more westward (for cases with half Saturn's radius).
At constant $\beta$ and water abundance, increasing the planetary radius makes the
flow more {\it eastward} for our 5-times-solar-water cases but either has minimal
effect or makes the equatorial flow more {\it westward} for our cases with
20-times-solar water.

In some cases, these dependences on radius and rotation rate can make the
difference between whether the equatorial flow is eastward or westward.  For
example, the Saturn cases with 5-times-solar water all transition from westward
to eastward equatorial flow as the rotation rate increases.  Likewise, cases with 
5-times-solar water and $\beta$ of 5--$10\times10^{-12}\rm\,m\,sec^{-1}$ exhibit
equatorial superrotation when performed at Saturn's nominal radius but equatorial
subrotation when performed at half Saturn's radius.

To summarize our simulations, equatorial superrotation occurs only at 
intermediate water abundances; large deep water abundance instead promotes 
the development of equatorial 
subrotation for all cases explored.  Everything else being equal,
greater rotation rate and planetary radius promote equatorial superrotation
when the water abundance is intermediate ($\sim5$-times solar).  The trends are
less clear at high water abundance ($\sim20$ times solar), but regardless
of the details our high-water-abundance cases all subrotate.  Taken together, the
sensitivity studies described here show that the equatorial superrotation in our
successful Jupiter and Saturn simulations (e.g., as shown in 
Fig.~\ref{jupiter_neptune_u_3d}) is enabled {\it not only} by the low-to-intermediate
water abundance adopted in those cases but {\it also} by the large planetary
radius and faster rotation rates of Jupiter and Saturn relative to Uranus
and Neptune.  Loss of any one of those factors would promote weaker 
superrotation or even a transition to equatorial subrotation.  This is
consistent with the fact that the Uranus/Neptune cases shown in 
Fig.~\ref{Neptune_U_cs} exhibit only rather weak superrotation (relative
to the Jupiter/Saturn cases) even when the water abundance is 1, 3, or 10
times solar.

\subsection{What drives the jets in weather layer}

\begin{figure}
 \centering
\includegraphics[width=5.5in]{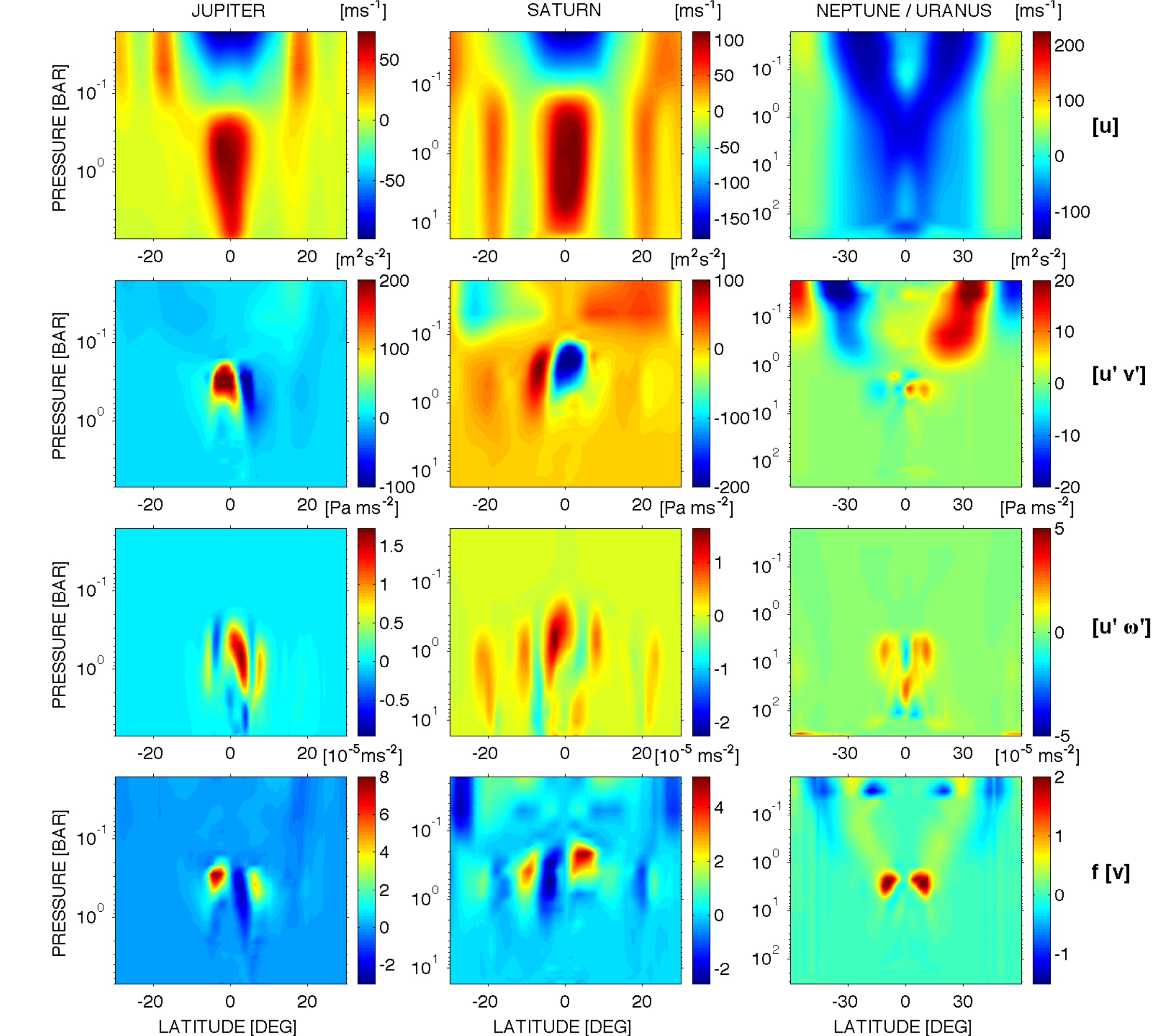}
\caption{Eddy fluxes and accelerations for Jupiter simulation with 3 times solar water
(left column), Saturn simulation with 5 times solar water (middle column), and
Uranus/Neptune simulation with 30 times solar water (right column).  Top row: zonal-mean
zonal winds.  Second row:  horizontal flux of eastward
momentum, $[ {\overline{u^\prime v^\prime}} ] + [ {\overline{u}}^* {\overline{v}}^* ]$.
Positive means northward flux and negative means southward flux.
Third row: vertical flux of eastward momentum $[ {\overline{u^\prime \omega^\prime}} ] + 
[ {\overline{u}}^* {\overline{\omega}}^* ]$.  Positive means downward flux and
negative means upward flux. Bottom row: Coriolis acceleration on
the mean-meridional circulation, $f [{\overline{v}}]$.  All quantities are averaged over
231 Earth days ending at day 578, 1600, and 578 Earth days, respectively, for the 
Jupiter, Saturn, and Uranus/Neptune cases.}
\label{winds_acc}
\end{figure}

Our Jupiter, Saturn, and Uranus/Neptune simulations show that the zonal jets
are predominantly driven by eddy and Coriolis acceleration.  
We use the total acceleration of the terms
on the right side of Eq.~\ref{eulerian-mean-mom} 
to characterize the main driving forces on the zonal flow, focusing here
on the mid- and low-latitude regions.
Figure~\ref{winds_acc} shows the zonal-mean zonal wind (top row),
horizontal eddy-momentum flux 
$[ {\overline{u^\prime v^\prime}} ] + [{\overline{u}}^* {\overline{v}}^* ]$
(second row), the vertical eddy-momentum flux $[ {\overline{u^\prime
\omega^\prime}} ] + [ {\overline{u}}^* {\overline{\omega}}^* ]$
(third row), and Coriolis acceleration $f [\overline{v}]$ (bottom row).
Three cases are shown --- our 3-times-solar water Jupiter case (left column),
5-times-solar Saturn case (middle row), and 30-times-solar Uranus/Neptune
case (right column).  The Jupiter and Saturn cases exhibits equatorial
superrotation while the Uranus/Neptune case exhibits equatorial subrotation.

Figure~\ref{winds_acc} shows that the horizontal eddy terms, vertical eddy terms, and Coriolis
accelerations all play important roles in the maintenance of the zonal winds.
In particular, for our Jupiter and Saturn cases, there are strong equatorward
fluxes of eastward momentum at pressures of $\sim0.2$--$1\,$bar and latitudes
of $\sim10^{\circ}$S to $10^{\circ}$N.  These imply an eastward acceleration
that helps to maintain the superrotating equatorial jet.  They additionally cause a 
{\it westward} acceleration at latitudes of $\sim10^{\circ}$N and S, where a divergence
in horizontal flux occurs.  In contrast, our Uranus/Neptune
case exhibits strong poleward fluxes of eastward momentum  at pressures 
$<1\,$bar and latitudes equatorward of $\sim40^{\circ}$.  These fluxes induce
westward equatorial acceleration, which maintains the strong westward equatorial flows
at low pressure.  A weaker version of the same phenomenon occurs in the Jupiter
and Saturn cases, which helps explain the westward equatorial stratospheric flow near the
top of the model (pressures
$<0.1\,$bar) in those cases.  Nevertheless, the Uranus/Neptune
case also shows a localized region (from $1$--$2\,$bars and latitudes $\sim10^{\circ}$S
to $10^{\circ}$N) where eastward momentum fluxes (albeit weakly) toward the equator, leading
to an eastward equatorial acceleration.  This relates to the fact that the westward jet,
once formed, weakens slightly between $\sim100$ and $1000\,$days (compare
second and third rows of Fig.~\ref{neptune_30s_evol}).  Interestingly, all three cases
also show an overall downward eddy flux of eastward momentum in the equatorial
regions underlying the region of horizontal eddy fluxes.  In the Jupiter/Saturn
cases, this term causes an acceleration counteracting that associated with
horizontal eddy-flux convergence and helps to explain why the eastward jets
penetrate to pressures $>10\,$bars despite the fact that the horizontal
eddy fluxes are confined primarily to pressures $<1\,$bar.  The Coriolis
accelerations (bottom row) show localized regions of eastward acceleration
centered just off the equator in all three cases (at pressures 0.2--$1\,$bar
for Jupiter/Saturn and $\sim2$--$8\,$bars for Uranus/Neptune) resulting 
from the effects of a meridional circulation cell near the equator.  In
all three cases, this acts to counteract a {\it westward} acceleration associated
with horizontal eddy flux convergence at the same location.

An estimate of magnitudes shows that all these acceleration terms are important.
For example, focusing on Jupiter and Saturn,
the Coriolis acceleration reaches peak values up to a few $\times10^{-5}\rm\,m\,s^{-2}$
(Fig.~\ref{winds_acc}, bottom row).
The acceleration caused by horizontal eddy flux convergence is minus the gradient of 
$[ {\overline{u^\prime v^\prime}} ] + [{\overline{u}}^* {\overline{v}}^* ]$,
which is approximately the difference in this quantity over a relevant length scale.
Figure~\ref{winds_acc}, second row, shows that the peak difference 
in $[ {\overline{u^\prime v^\prime}} ] + [{\overline{u}}^* {\overline{v}}^* ]$ is
$\sim200\rm\,m^2\,s^{-2}$ and occurs over a latitudinal length scale of
$\sim8000\,$km, implying an acceleration of $\sim3\times10^{-5}\rm\,m\,s^{-2}$.
Likewise from Figure~\ref{winds_acc}, third row, the peak difference in  $[ {\overline{u^\prime
\omega^\prime}} ] + [ {\overline{u}}^* {\overline{\omega}}^* ]$ is
$\sim1\rm\,Pa\,m\,s^{-2}$, which occurs over a pressure interval of $\sim10^5\,$Pa,
implying an acceleration of $\sim10^{-5}\rm\,m\,s^{-2}$.

For all these cases, there are partial cancellations between the
individual terms that generally leads to a net acceleration smaller
than the magnitude of the dominant individual terms.  More than $\sim5$--$10^{\circ}$
away from the equator, there is an anticorrelation between the 
Coriolis acceleration and acceleration due to horizontal eddy convergence, 
leading to a significant cancellation between these terms.  
Accelerations due to vertical eddy convergence play only a small
role in these regions because the ratio of accelerations due to
vertical and horizontal eddy convergences tends to scale as 
the Rossby number, which is small away from the equator.  However,
near the equator the Coriolis acceleration is weak, and 
within $\sim3$--$5^{\circ}$ latitude of the equator, the dominant cancellation is between
the horizontal and vertical eddy terms for Jupiter, Saturn and
Uranus/Neptune.  Because of these partial cancellations, the net
acceleration is relatively small once the jets have spun up,
leading to only gradual change in the zonal-jet speeds over time.

\subsection{Comparison between simulations and observations}

\begin{figure}
 \centering
\includegraphics[width=5in]{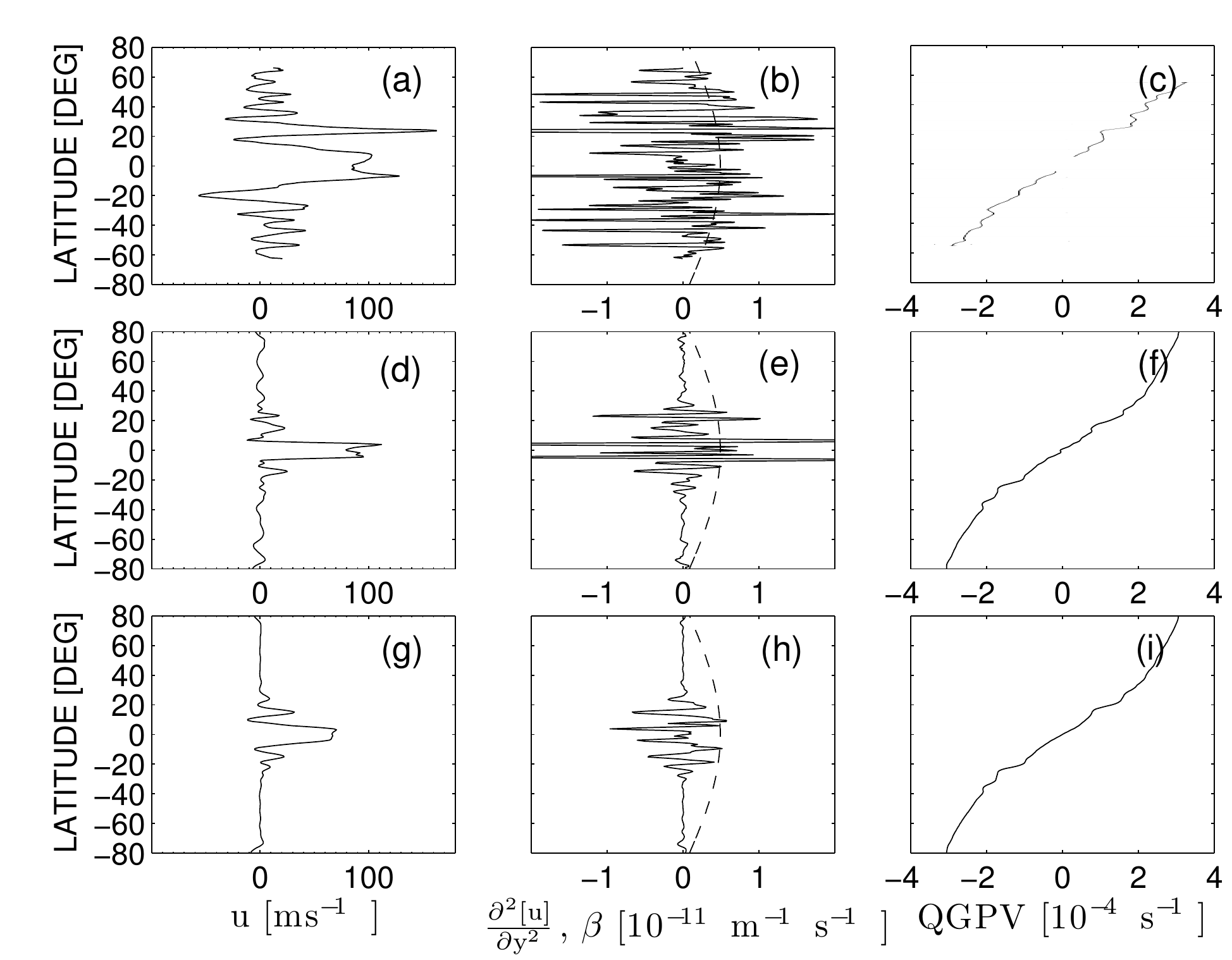}
\caption{Zonal-mean zonal winds (left column), curvature of
zonal-mean-zonal winds $\partial^2 [u]/\partial y^2$ with northward
distance $y$ (middle column) at $\sim0.6\,$bars, and zonal-mean quasigeostrophic potential vorticity
(right column) at $\sim0.12\,$bars.  In the middle column,
$\beta$ is included as a dashed curve for comparison.
Top row (a, b, and c) shows Voyager 2 measurements; winds are
from \citet{1986Icar...65..335L} and PV is from \citet{Read2006a}.
Middle row (d, e, and f) shows the zonal winds properties at $163^\circ$ east in our Jupiter-type simulation
with 3 times solar water abundance 
at 2200 Earth days.  Bottom row (g, h, and i) shows the zonal mean zonal winds properties in our
Jupiter simulation at 2200 Earth days.}
\label{Jupiter_sim_observ}
\end{figure}

Now we compare our simulated jet profiles to the observed jet profiles 
and their stability.   Jupiter and Saturn's cloud-level winds violate 
the barotropic stability criterion 
\citep{ingersoll-etal-1981}
\begin{equation}
\frac{\partial^2 [u]}{\partial{y^2}} < \beta.
\end{equation}
The observed winds also violate the Charney-Stern criterion which states that 
jets are stable if their potential vorticity profile is monotonic in latitude 
\citep{dowling-1995}.  
Figure~\ref{Jupiter_sim_observ} shows the comparison between observed 
and simulated jet profiles for Jupiter. 
The top row shows observations. The middle row 
shows simulated winds at $\rm 163^\circ$ longitude. The bottom row 
shows zonal-mean simulated zonal winds. The wind profile shown is at 
$\rm \sim 0.6\,bars$ and potential vorticity shown is at $\rm \sim 0.12\,bars$. 
We calculate quasi-geostrophic potential vorticity $q_G$ following 
\citet{Read2006a}:
\begin{equation}
q_G=f+\zeta -f{\partial\over\partial p} \left[{p\Delta T(\lambda,\phi,p)
\over s(p)T_a(p)}\right],
\label{qgpv}
\end{equation}
where $\zeta$ is relative vorticity calculated on isobars,
$T_a(p)=\langle{T(\lambda,\phi,p)}\rangle$ is the horizontal mean
temperature calculated on isobars, $\Delta{T(\lambda,\phi,p)}=
T(\lambda,\phi,p)-T_a(p)$ is deviation of
temperature from its horizontal mean, and $s(p)$ is a stability factor
defined as
\begin{equation}
s(p)=-{\partial{\langle{\ln(\theta)}\rangle}\over{\partial{\ln p}}}.
\label{s-fac}
\end{equation}
where $\langle\theta\rangle$ is the horizontal mean potential temperature
calculated on isobars.

Our Jupiter simulation with 3 times solar water 
abundance shows that the zonal winds are slightly weaker and 
the equatorial superrotation is narrower than observed.  The zonal winds 
in some longitudinal cross sections violate the  barotropic
and Charney-Stern stability criteria. Figure~\ref{Jupiter_sim_observ}(e) 
shows that the wind curvature 
$\partial^2 [u] /{\partial{y^2}}$ in a longitudinal slice 
(here arbitrarily chosen as $\rm 163^\circ$ east) exceeds $3\beta$ at some latitudes. 
At the same time, 
the latitudinal gradient of quasigeostrophic potential vorticity changes sign at 
many places, as shown in Fig.~\ref{Jupiter_sim_observ}(f). The zonal-mean 
zonal wind, however,  has a much weaker violation of the barotropic 
stability criterion. It only exceeds $\beta$ at $\rm \pm 20^\circ$ in 
latitude. 
 
\begin{figure}
 \centering
\includegraphics[width=5in]{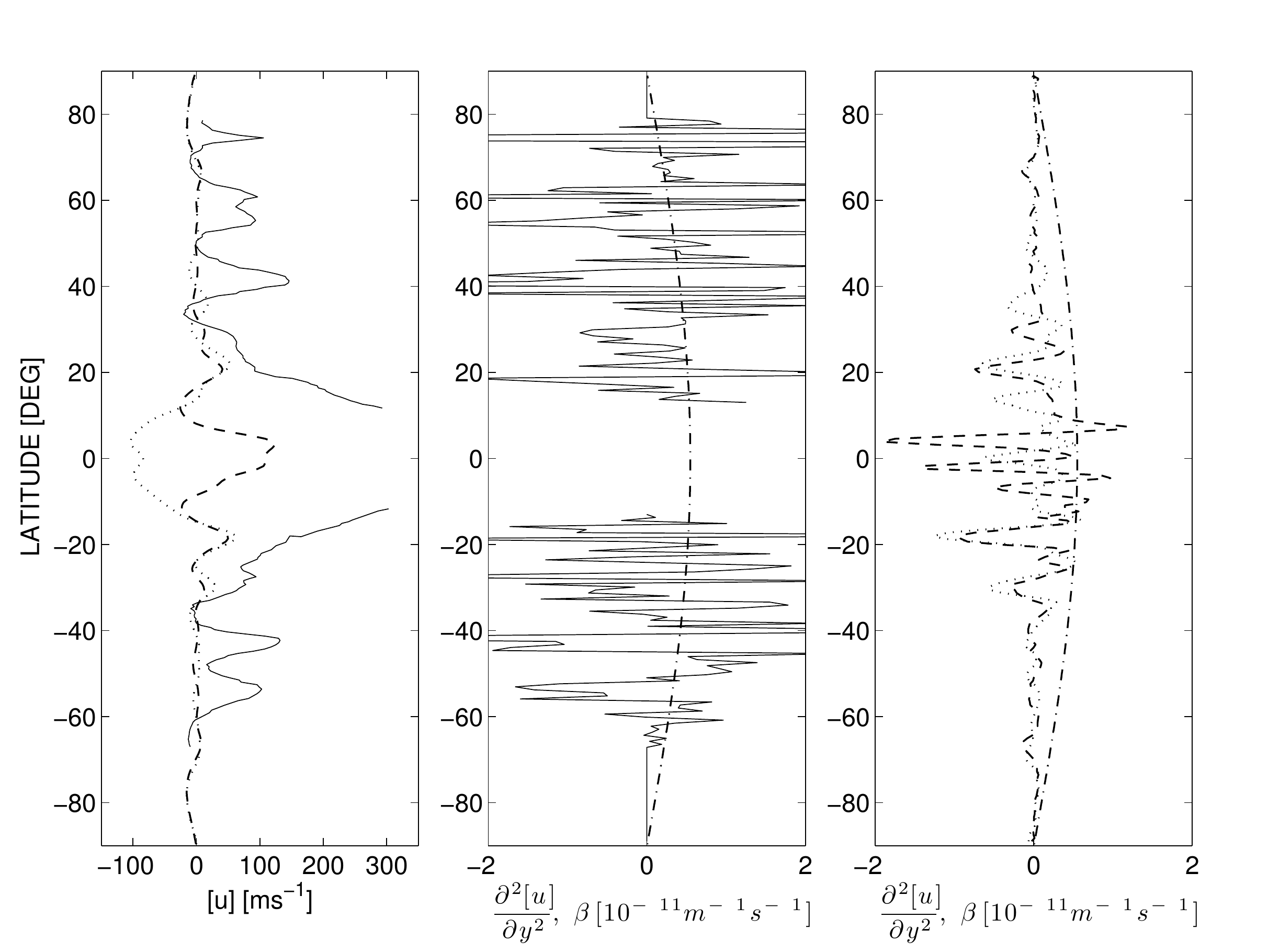}
\caption{Saturn zonal-mean zonal winds (left), curvature of observed
zonal-mean-zonal winds $\partial^2 [u]/\partial y^2$ with northward
distance $y$ (middle) and curvature  $\partial^2 [u]/\partial y^2$
of our Saturn simulations (right).  In all panels,
solid curve is Saturn observations from Voyager, 
dashed curve is Saturn simulation with 5 times solar water, and 
dotted curve is Saturn simulation with 10 times solar water.  
In the middle and right panels, $\beta$ is included as a dot-dashed 
curve for comparison.  Simulation results are shown at 1 bar and
observations are also roughly at 1 bar.}
\label{Saturn_sim_observ}
\end{figure}

Figure~\ref{Saturn_sim_observ} compares observations and simulations
for Saturn.  Zonal-mean zonal wind profiles are shown on the left,
with Voyager observations (solid), our Saturn 5-times-solar-water
simulation (dashed), and our Saturn 10-times-solar-water simulation
(dotted).  The middle and right panels show $\partial^2[u]/\partial y^2$
for the observations and our simulations, respectively.  As can
be seen, our simulations produce winds with speeds smaller than
observed, especially in the equatorial jet.  Interestingly, our
Saturn simulation with 5-times-solar water, which produces 
equatorial superrotation of $150\rm\,m\,s^{-1}$, violates
the barotropic stability criterion at some latitudes.  In
contrast, our Saturn simulation with 10-times-solar water, which
produces equatorial subrotation, satisfies the barotropic stability
criterion at all latitudes.

\begin{figure}
 \centering
\includegraphics[width=5in]{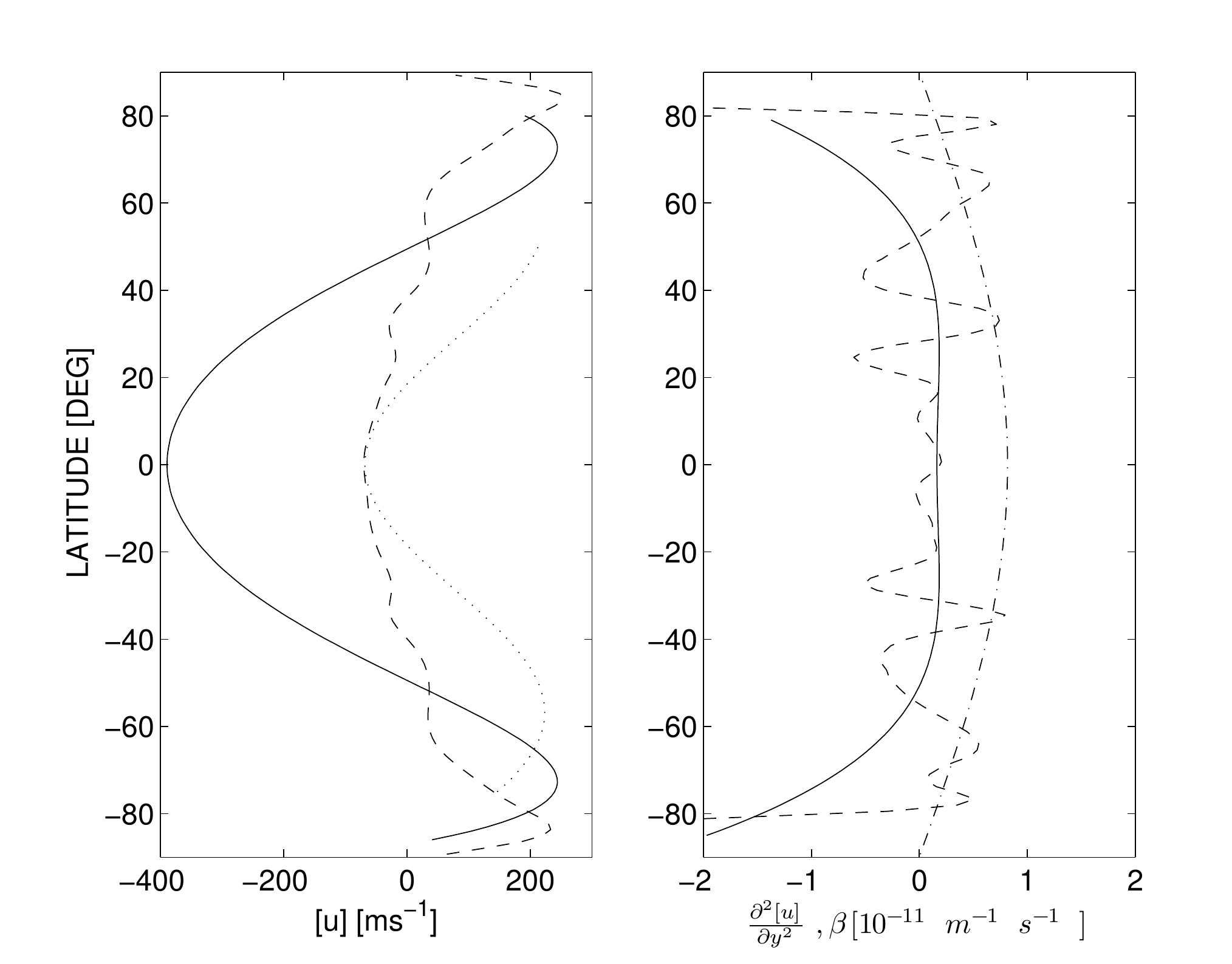}
\caption{Zonal-mean zonal winds (left column) and curvature of
zonal-mean-zonal winds $\partial^2 [u]/\partial y^2$ with northward
distance $y$ (right column) at $\sim 1\,$bar.  In the right column,
$\beta$ is included as a dot-dashed curve for comparison.
The solid line shows Voyager 2 measurements of Neptune winds; winds are
from \citet{Sromovsky1993}. The dashed line shows the zonal winds 
properties in our Neptune-type simulation with 30 times solar water abundance
 at 2200 Earth days. The dotted line shows HST measurement of 
Uranus winds; winds are from \citet{Hammel2001}.}
\label{Neptune_sim_observ}
\end{figure}

We also compare jet profiles in our Uranus/Neptune simulation with 30 times 
solar water abundance to observations. Figure~\ref{Neptune_sim_observ} shows 
the observed winds on Uranus and Neptune and simulated winds at $\rm \sim 1\,$ bar. 
Our simulated zonal winds share similarities with the observed winds, especially 
the three-jet feature with a subrotating equator and high-latitude eastward jets.
Comparing to the zonal winds on Neptune, the equatorial subrotation 
in our simulation achieves speed of
$\rm 100\,m\,s^{-1}$ which is weaker than the observed $\rm 400\, m\,s^{-1}$; 
the polar eastward winds, though appearing at too high a latitude, 
have similar strength 
as observed. On the other hand, the simulated jets have similar strength 
as both westward and eastward jets on Uranus. However, the westward jet on Uranus 
is much narrower than that in our simulation and on Neptune. Furthermore, Uranus 
has eastward jets peaking in the mid-latitudes rather than near the poles
as in our simulation. 
Interestingly, zonal winds in our simulation violate the barotropic 
stability criterion while those in observations do not. 

Even though the zonal jets in our simulations violate the barotropic 
stability criterion, the time-averaged jets are still stable which suggests a stable 
configuration between the eddies and zonal jets. This stable configuration also 
exists in our previous simulations \citep{lian-showman-2008}.

\subsection{Morphology of eddies generated by large-scale latent heating}
\label{storms}
In our Jupiter and Saturn simulations, large-scale latent
heating and rising motion often become concentrated into
localized regions, leading to the development of small
($\sim$3000--10,000-km diameter) warm-core eddies that play a key role in
driving the flow.  These events share similarities with 
storm clouds observed on Jupiter and Saturn, so here we describe them in detail.  
For lack of a better word, we call these events ``storms'' but remind the
reader that our model lacks non-hydrostatic motions and sub-grid-scale
parameterizations of true moist convection.

\begin{figure}
 \centering
\includegraphics[width=5in]{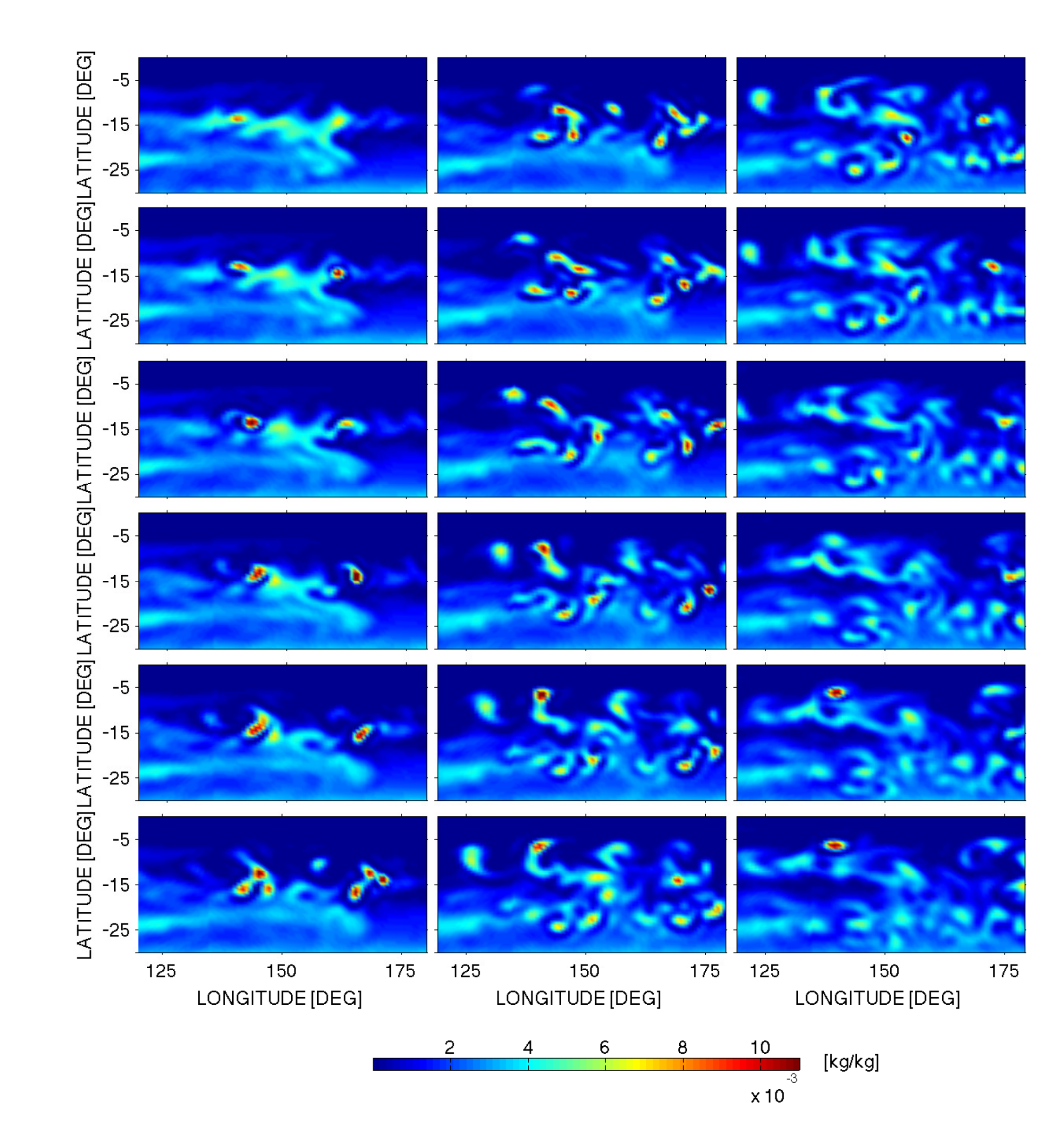}
\caption{A zoom-in showing evolution of water-vapor mixing ratio 
at the 5-bar level in a portion of the domain centered around
a moist-convection event
for our Jupiter simulation with 3 times the solar water abundance. 
Left column (from top down) starts from 1185 Earth days and ends at 1199 Earth days. 
Middle column (from top down) starts from 1201.8 Earth days and ends at  1215.8 Earth days.
Right column (from top down) starts from 1218.6 Earth days and ends at 1232.6 Earth days. The time 
interval between two adjacent time frames in a column is 2.8 Earth days.}
\label{jupiter_S_slides}
\end{figure}

\begin{figure}
 \centering
\includegraphics[width=5in]{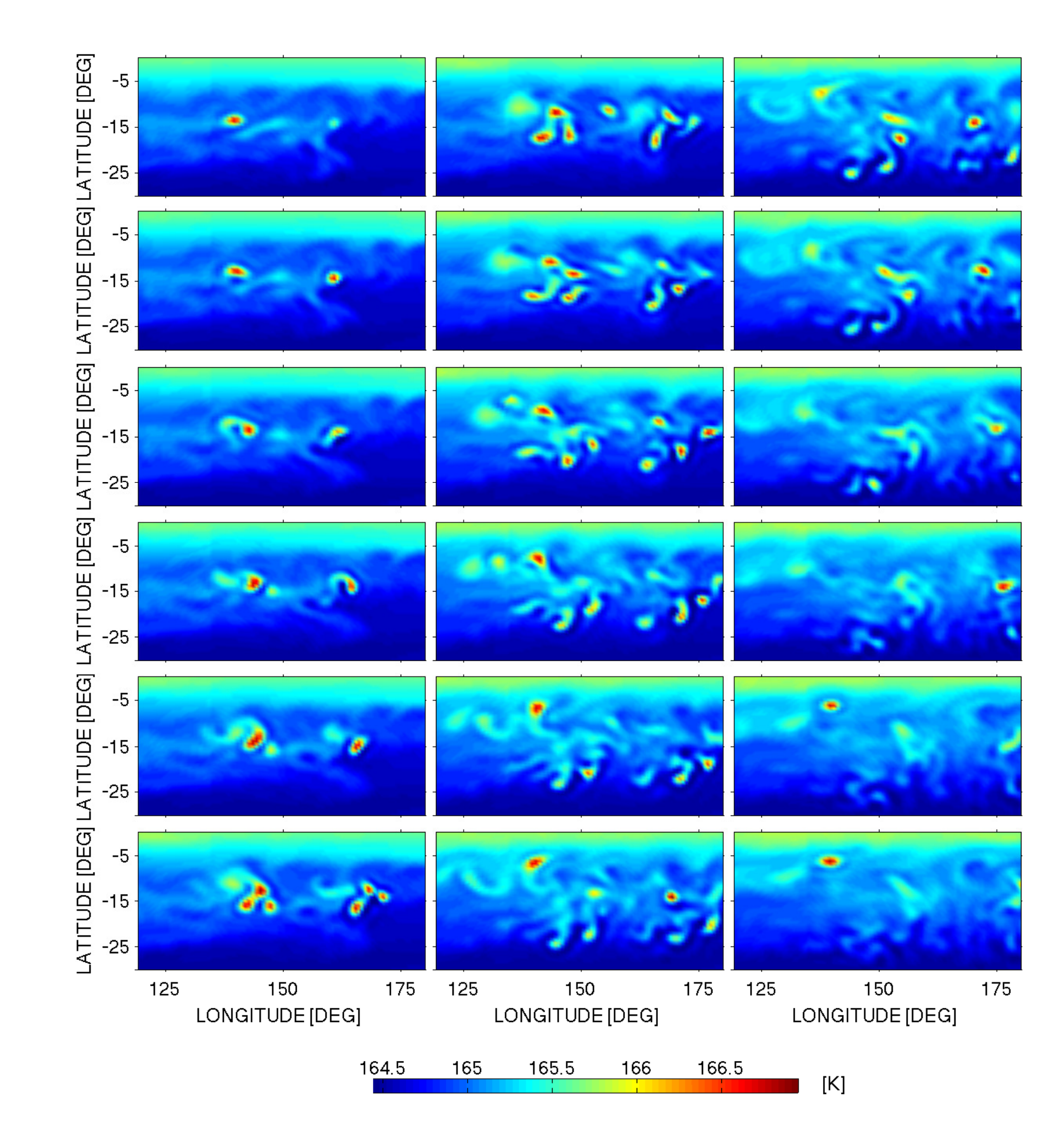}
\caption{Same as Fig.~\ref{jupiter_S_slides} but shows potential temperature
at the 5-bar level.}
\label{jupiter_T_slides}
\end{figure}

\begin{figure}
 \centering
\includegraphics[width=5in]{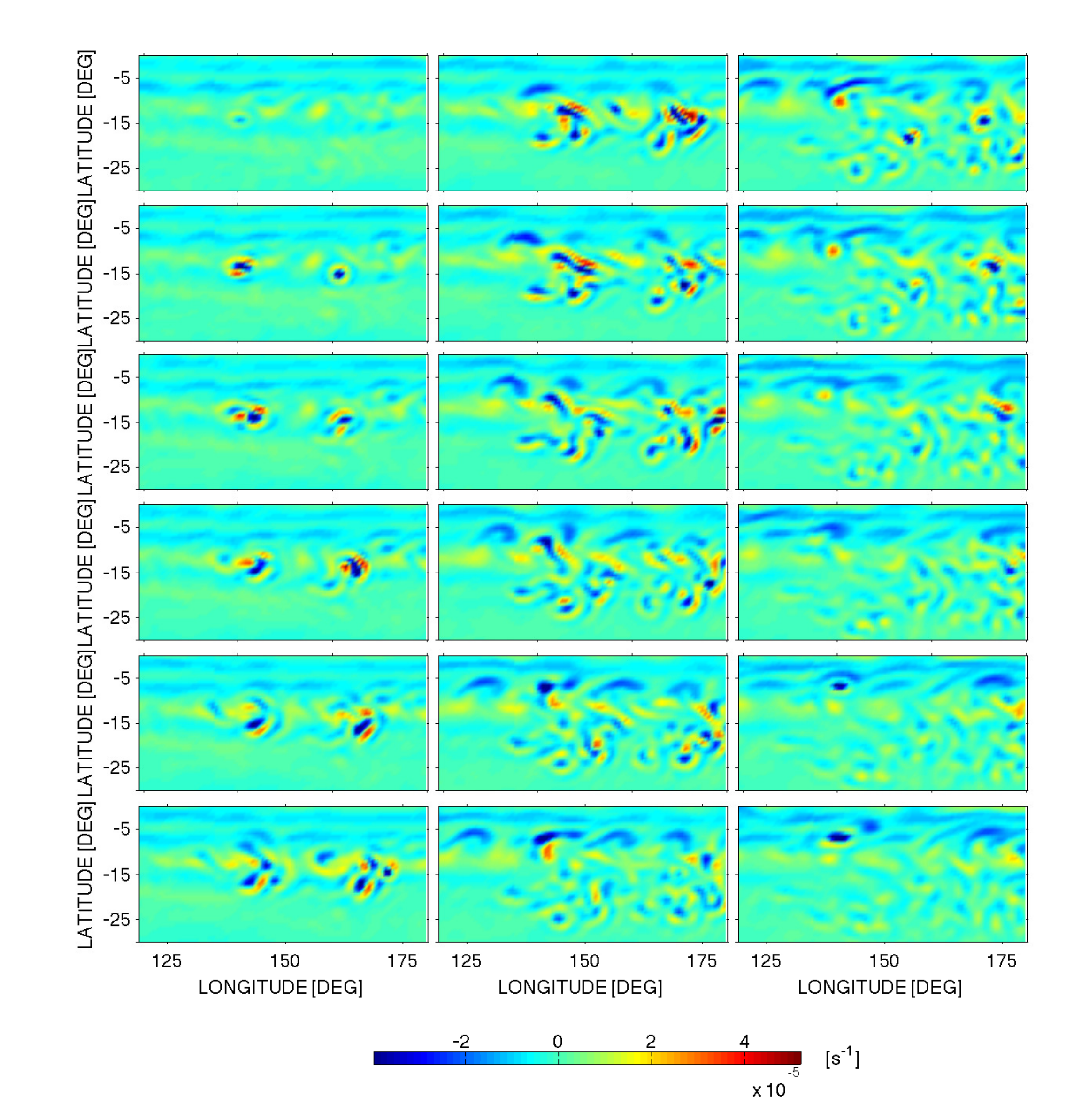}
\caption{Same as Fig.~\ref{jupiter_S_slides} but shows relative vorticity
at the 5-bar level.  Blue is cyclonic and red is anticyclonic.}
\label{jupiter_vort_r_slides}
\end{figure}

\begin{figure}
 \centering
\includegraphics[width=5in]{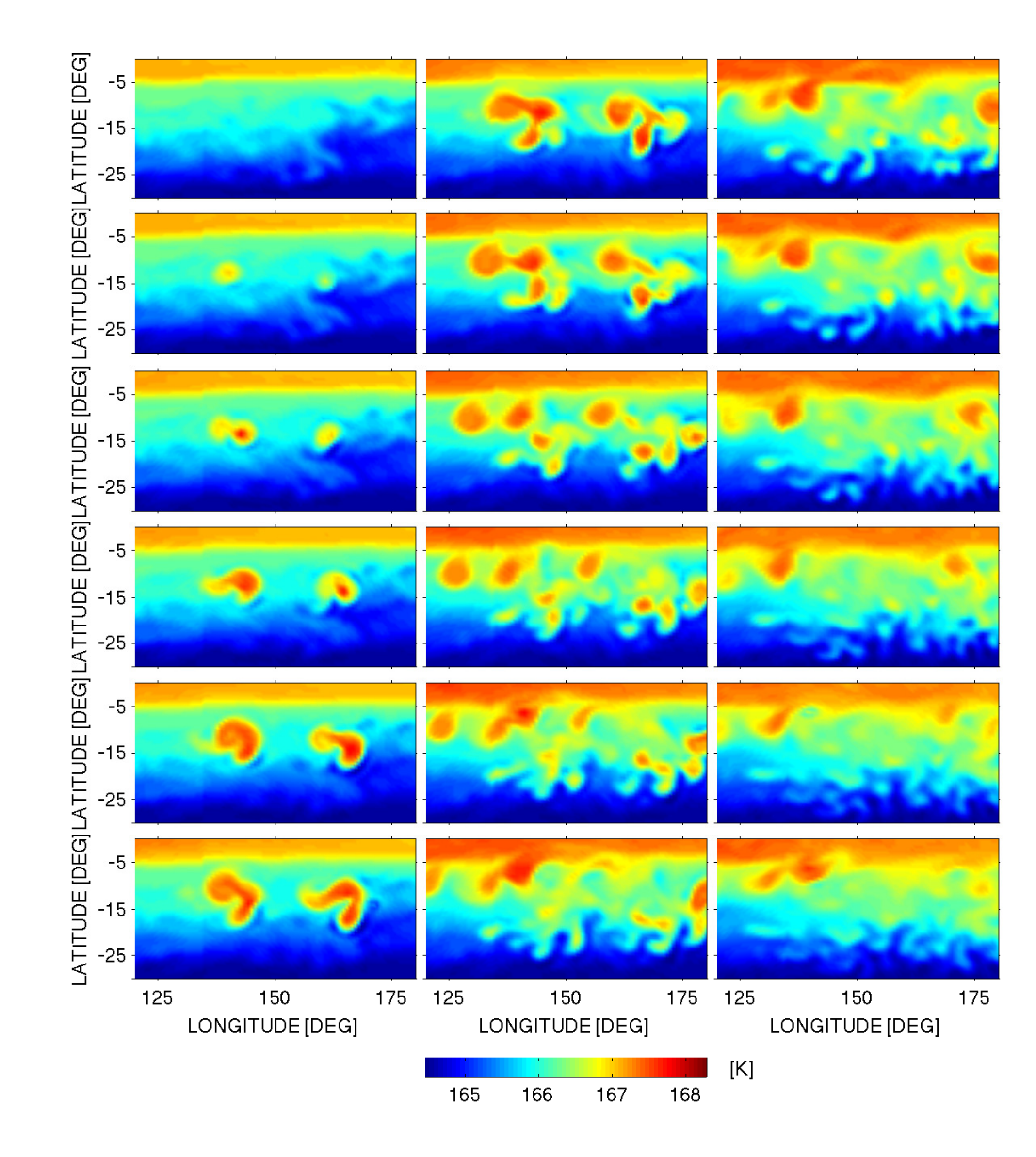}
\caption{Same as Fig.~\ref{jupiter_S_slides} but shows potential temperature
at the 0.9-bar level.}
\label{jupiter_T_slides_1bar}
\end{figure}

Figures~\ref{jupiter_S_slides}--\ref{jupiter_T_slides_1bar} depict
time sequences that zoom into the region around one such event;
the sequences start at the upper left and move first downward and then
right in 2.8-Earth-day intervals.    
Figure~\ref{jupiter_S_slides}, which depicts 
the water-vapor mixing ratio, shows the active storms as bright
red regions with much greater water abundance than the surroundings.
Figure~\ref{jupiter_T_slides} demonstrates that these regions of
high humidity are warmer than the surroundings, which is the direct
result of latent heating in the rising air.  Plots of vertical velocity
(not shown) show that these warm, moist regions are ascending.
Figure~\ref{jupiter_vort_r_slides} depicts the relative vorticity;
a careful comparison with Figs.~\ref{jupiter_S_slides}--\ref{jupiter_T_slides} shows that, at the base of
the storms near 5 bars, the hot, moist ascending regions generally
exhibit cyclonic relative vorticity, which results from the 
Coriolis acceleration on the air converging horizontally into the 
base of the storm at that pressure.  Although the storms are
extremely dynamic, they are self-generating and can last for tens
of days.

Interestingly, the cyclonic
regions at the base of the updrafts 
(blue in Fig.~\ref{jupiter_vort_r_slides})
typically co-exist with one or more localized anticyclonic
regions (red in Fig.~\ref{jupiter_vort_r_slides}), which are
locations where air descends, horizontally diverges, and thus
spins up anticyclonically.  The existence of descending regions
in proximity to the ascending storm centers results from mass
continuity; as air rises in an active storm center, continuity
demands descent in the surrounding environment. A similar phenomenon
occurs around storms on Earth; theoretical
solutions show that such descent is typically confined to
regions within a deformation radius of the ascending storm center
\citep[][pp. 329-333]{Emanuel_1994}.  This result can explain the 
close proximity of the ascending and descending regions in our simulations,
as well as the fact that Jovian thunderstorms
are often observed next to localized regions that are clear down
to the 4-bar level or deeper \citep{1998Icar..135..230B, 
2000Natur.403..628G}.

The behavior near the tops of the simulated storms differs 
significantly from that near their bases.  This is illustrated in 
Fig.~\ref{jupiter_T_slides_1bar}, which shows the temperature
at 0.9 bars for the same storm event depicted in 
Figs.~\ref{jupiter_S_slides}--\ref{jupiter_vort_r_slides}.  Hot
regions in Fig.~\ref{jupiter_T_slides} generally correlate with
hot regions in Fig.~\ref{jupiter_T_slides_1bar}, which results from
the fact that the hot, moist air at 5 bars generally continues rising
until reaching altitudes at and above the 1-bar level.  However, 
the localized hot regions at 0.9 bars are much larger than at 5 bars.  
This presumably results from the horizontal divergence that occurs near 
the storm top, which spreads the hot regions out into ``anvils'' whose
horizontal extent significantly exceeds that at the base of the
storm.  Conversely, we speculate that the horizontal convergence at the 
storm base helps to keep the hot regions horizontally confined at that 
pressure. The horizontal divergence near the storm top also implies that
the storms generate regions of anticyclonic vorticity as seen
at 1 bar (not shown).  This is consistent with observations of
Jovian thunderstorms, which generally also develop anticyclonic
vorticity at the ammonia cloud level \citep{2000Natur.403..628G}.

The simulated storms exhibit a complex evolution.   The storms
often appear in clusters with several simultaneously active
centers (Figs.~\ref{jupiter_S_slides}--\ref{jupiter_T_slides_1bar}); 
this could occur because the large-scale environment
in that region is primed for storm generation, but it also
suggests that the storms may self-interact in a way that can trigger
new storms.  As shown in Figure~\ref{jupiter_T_slides_1bar},
active storm centers sometimes shed warm-core vortices that
have lifetimes up to tens of
days and propagate downstream away from the active storm region.  
Fig.~\ref{jupiter_T_slides} exhibits several 
examples where features appear to ``die out'' at 5 bars, but a comparison with 
Fig.\ref{jupiter_T_slides_1bar} shows that, in several cases,
the features have not decayed but instead have simply ascended
to the $\sim1$-bar level.  Indeed, several of the warm-core
vortices visible in Fig.~\ref{jupiter_T_slides_1bar} have
essentially no signature at 5 bars, indicating that these
vortices are no longer fed by upward motion
from near the water-condensation level.

Importantly, the properties of our simulated storms --- including
their size, amplitude, lifetime, and morphology --- are 
self-consistently generated by the 
dynamics and are therefore predictions of our model.  
This contrasts with previous studies attempting to model the effect of 
moist convection on the large-scale flow \citep{Li-2006, showman-2007},
which introduced mass pulses by hand to represent storms, and which 
imposed the size, shape, amplitude, and lifetime of the storms
as free parameters.

Despite the lack of small-scale convective dynamics in our model,
our simulated storms show an encouraging resemblance to the evolution
of cloud morphology in real
moist-convection events on Jupiter and Saturn.  On Jupiter,
individual storm clouds often last for up to several days and expand to
diameters up to $\sim1000$--$3000\,$km \citep{2000Natur.403..628G, 
Porco2003, sanchez-lavega-etal-2008}, although rare events sometimes 
reach sizes of $10,000\,$km or more \citep{hueso-etal-2002} and lifetimes
up to 10 days \citep{Porco2003}.  On Saturn, storms lasting
tens of days have been observed by Voyager and Cassini
\citep{Sromovsky1983, Porco2005, dyudina-etal-2007}; individual
active storm centers reach $\sim2000\,$km diameters within
an active storm complex up to $\sim6000\,$km across.   The observed
clouds are probably large-scale anvils fed by numerous individual
convective storms that become spatially organized, analogous to ``mesoscale
convective complexes'' on Earth \citep{2000Natur.403..628G}.  
As already described, the rapid expansion and
generation of anticyclonic vorticity near the tops of our simulated
storms and the close proximity of ascending active storm centers to subsiding 
regions are consistent with observations of jovian storms, although
our storm sizes and lifetimes are somewhat too large.
The creation of (generally short-lived) vortices from latent 
heating, as occurs in our simulations, has
not been clearly observed on Jupiter, but storm-generated vortices
have tentatively been
captured in Cassini images on Saturn \citep{Porco2005, dyudina-etal-2007}.
A similar phenomenon was obtained in the shallow-water
simulations of \citet{showman-2007}.

It is worth re-iterating that our simulations adopt local hydrostatic
balance and thus cannot capture the strong vertical accelerations
associated with convectively unstable motions and lightning generation.  
At present, such phenomena can only be resolved by regional-scale
non-hydrostatic models \citep[e.g.][]{yair-etal-1995, hueso-etal-2001},
although the effects of such convective motions 
on the large-scale flow could potentially be represented in large-scale 
models using a cumulus parameterization.
The large horizontal dimensions of observed jovian/saturnian 
storm anvil clouds, and the qualitative similarity of our results to the 
observed storms, supports the possibility that
such efforts could prove fruitful in attempting to explain
observations of storm-cloud evolution on Jupiter and Saturn.

\section{Conclusion}

We presented global, three-dimensional numerical models to simulate the 
formation of zonal jets by large-scale latent heating on the giant planets.
These models explicitly include water vapor and its 
condensation and the resulting latent heating. We find that latent heating 
can naturally produce banded zonal jets similar to those observed on
the giant planets. Our Jupiter and Saturn 
simulations develop $\rm \sim 20$ zonal jets and Uranus/Neptune simulations 
develop $\rm 3 - 7$ zonal jets depending on the water abundance. The jet spacing is consistent  
with Rhines scale $\pi (2U/\beta)^{1/2}$. The zonal jets in our 
simulations produce
modest violations of the barotropic and Charney-Stern stability criteria at some latitudes. 

In our simulations,
condensation of water vapor releases latent heat and produces baroclinic eddies.  These
eddies interact with the large-scale flow and the $\beta$ effect to pump momentum
up-gradient into the zonal jets.  At the same time, a meridional circulation develops
whose Coriolis acceleration counteracts the eddy accelerations in the weather layer.
This near-cancellation of eddy and Coriolis accelerations
leads to slow evolution of the zonal jets and maintains the steadiness 
of zonal jets in the presence of continual forcing.
Such a process was suggested by \citet{2006Icar..182..513S} and \citet{Del_Genio2007}
and occurred also in the simulations of \citet{showman-2007} and \citet{lian-showman-2008}.

Our simulations also produce equatorial superrotation for Jupiter and Saturn and subrotation for Uranus
and Neptune.  Although a number of previous attempts have been made to produce superrotation
on Jupiter/Saturn and subrotation on Uranus/Neptune, previous models have generally lacked
an ability to produce {\it both} superrotation on Jupiter/Saturn and subrotation on Uranus/Neptune
without introducing {\it ad hoc} forcing or tuning of model parameters.
Ours is the first study to naturally produce such dual behavior, without tuning, 
within the context of a single model.  
Although the speeds of the superrotation and subrotation are weaker than observed, 
our simulations provide a possible mechanism to explain the dichotomy in equatorial-jet
direction between the gas giants (Jupiter/Saturn) and ice giants (Uranus/Neptune) within
the context of a single model. In our simulations, the strength, 
scale, and direction of the equatorial flow are strongly affected by the abundance of water vapor 
as well as by the planetary radius and rotation rate.  
Equatorial superrotation preferably forms in simulations with modest water vapor abundance
and is further promoted by large planetary radii and fast rotation rates, as occur
at Jupiter and Saturn.  In contrast, high water abundance leads to equatorial subrotation 
regardless of the planetary radius and rotation rate explored here.  In this
picture, the dichotomy in the equatorial jet direction between the gas and ice giants
would result from a combination of the faster rotation rates, larger radii, and
probable lower water abundances on Jupiter/Saturn relative to those on Uranus/Neptune.

Despite this encouraging result, Saturn poses a possible difficulty with this 
picture.  Our Saturn simulations generated equatorial superrotation when the 
water abundance is five times solar but equatorial subrotation when it is ten 
times solar.  Saturn's actual water abundance is unknown but could easily lie 
anywhere within this range.  Moreover, even our five-times-solar Saturn case produced 
a superrotation that is much weaker and narrower than observed --- about 
$120\rm \,m\,s^{-1}$ in the simulation versus $\sim400\rm \,m\,s^{-1}$ on the real 
planet.  (The equatorial jet in our Jupiter simulations is also too narrow, 
although the discrepancy is less severe.)  However, a 
variety of processes excluded here (e.g., cloud microphysics, realistic radiative 
transfer, and moist convection) could significantly affect the jet profile.  Future
investigations that include these effects are needed before a definite assessment can be 
made.

Consistent with our earlier work, we again find that our simulations generate deep barotropic 
jets despite the localization of the eddy accelerations to the weather layer (pressures less
than $\sim7\,$bars in our Jupiter simulations, for example). In some simulations,
the deep jets have similar strength as the winds in weather layer. The mechanism that forms 
the deep jets is similar to the mechanism we previously identified 
\citep{lian-showman-2008,2006Icar..182..513S}. However, 
these deep jets are affected by the redistribution of water vapor, which makes the diagnostics 
very complicated in the present case.

Our simulations successfully produce the large-scale dynamic features 
on Jupiter and Uranus/Neptune under the effect of large-scale latent heating.
However, our simulations lack long-lived vortices such as the Great Red Spot 
on Jupiter and Great Dark Spot on Neptune. We also ignore the precipitation 
and re-evaporation of condensates and use an idealized radiative cooling 
scheme. The grid resolution in our simulations is relatively low for 
resolving the mesoscale moist convection events which have 
typical horizontal scales of $1000$ kilometers \citep{Little1999, Porco2003}. 
Future models can include cloud physics, a sub-grid-scale
parameterization of cumulus convection,
a more realistic radiative transfer scheme, and explore the coupling 
between the deep interior and the weather-layer processes identified here.

\section{Acknowledgement}

This research was supported by NASA Planetary Atmospheres grant NNG06GF28G to APS.

\label{lastpage}

\bibliography{bibliography.bib}

\begin{thebibliography}{73}
\expandafter\ifx\csname natexlab\endcsname\relax\def\natexlab#1{#1}\fi
\expandafter\ifx\csname url\endcsname\relax
  \def\url#1{\texttt{#1}}\fi
\expandafter\ifx\csname urlprefix\endcsname\relax\def\urlprefix{URL }\fi

\bibitem[{{Arakawa}(2004)}]{arakawa-2004}
{Arakawa}, A., Jul. 2004. {The Cumulus Parameterization Problem: Past, Present,
  and Future.} Journal of Climate 17, 2493--2525.

\bibitem[{{Aurnou} et~al.(2007){Aurnou}, {Heimpel}, and
  {Wicht}}]{aurnou-etal-2007}
{Aurnou}, J., {Heimpel}, M., {Wicht}, J., 2007. {The effects of vigorous mixing
  in a convective model of zonal flow on the ice giants}. Icarus 190, 110--126.

\bibitem[{{Aurnou} and {Olson}(2001)}]{2001GeoRL..28.2557A}
{Aurnou}, J.~M., {Olson}, P.~L., 2001. {Strong zonal winds from thermal
  convection in a rotating spherical shell}. \grl 28, 2557--2560.

\bibitem[{{Baines} et~al.(1995){Baines}, {Hammel}, {Rages}, {Romani}, and
  {Samuelson}}]{baines-etal-1995}
{Baines}, K.~H., {Hammel}, H.~B., {Rages}, K.~A., {Romani}, P.~N., {Samuelson},
  R.~E., 1995. {Clouds and aerosols in the atmosphere of Neptune}. In Neptune
  and Triton (Ed. by D. P. Cruikshank, University of Arizona Press), 489--546.

\bibitem[{{Baines} et~al.(2009){Baines}, {Momary}, {Fletcher}, {Showman},
  {Roos-Serote}, {Brown}, {Buratti}, {Clark}, and
  {Nicholson}}]{baines-etal-2009}
{Baines}, K.~H., {Momary}, T.~W., {Fletcher}, L.~N., {Showman}, A.~P.,
  {Roos-Serote}, M., {Brown}, R.~H., {Buratti}, B.~J., {Clark}, R.~N.,
  {Nicholson}, P.~D., 2009. {Saturn's north polar cyclone and hexagon at depth
  revealed by Cassini/VIMS}. Planet. Space Sci. (submitted).

\bibitem[{{Banfield} et~al.(1998){Banfield}, {Gierasch}, {Bell}, {Ustinov},
  {Ingersoll}, {Vasavada}, {West}, and {Belton}}]{1998Icar..135..230B}
{Banfield}, D., {Gierasch}, P.~J., {Bell}, M., {Ustinov}, E., {Ingersoll},
  A.~P., {Vasavada}, A.~R., {West}, R.~A., {Belton}, M.~J.~S., 1998. {Jupiter's
  Cloud Structure from Galileo Imaging Data}. Icarus 135, 230--250.

\bibitem[{{Barcilon} and {Gierasch}(1970)}]{barcilon-gierasch-1970}
{Barcilon}, A., {Gierasch}, P., 1970. {A Moist, Hadley Cell Model for Jupiter's
  Cloud Bands.} Journal of Atmospheric Sciences 27, 550--560.

\bibitem[{{Borucki} and {Williams}(1986)}]{borucki-williams-1986}
{Borucki}, W.~J., {Williams}, M.~A., 1986. {Lightning in the Jovian water
  cloud}. J. Geophys. Res. 91, 9893--9903.

\bibitem[{{Campin} et~al.(2004){Campin}, {Adcroft}, {Hill}, and
  {Marshall}}]{campin2004}
{Campin}, J.-M., {Adcroft}, A., {Hill}, C., {Marshall}, J., 2004. {Conservation
  of properties in a free-surface model}. Ocean Modelling 6, 221--244.

\bibitem[{{Cho} and {Polvani}(1996{\natexlab{a}})}]{cho-polvani-1996a}
{Cho}, J.~Y.-K., {Polvani}, L.~M., 1996{\natexlab{a}}. The morphogenesis of
  bands and zonal winds in the atmospheres on the giant outer planets. Science
  8~(1), 1--12.

\bibitem[{{Cho} and {Polvani}(1996{\natexlab{b}})}]{1996PhFl....8.1531C}
{Cho}, J.~Y.-K., {Polvani}, L.~M., 1996{\natexlab{b}}. {The emergence of jets
  and vortices in freely evolving, shallow-water turbulence on a sphere}.
  Physics of Fluids 8, 1531--1552.

\bibitem[{{Christensen}(2001)}]{2001GeoRL..28.2553C}
{Christensen}, U.~R., 2001. {Zonal flow driven by deep convection in the major
  planets}. \grl 28, 2553--2556.

\bibitem[{{Del Genio} et~al.(2007){Del Genio}, {Babara}, {Ferrier},
  {Ingersoll}, {West}, {Vasavada}, {Spitale}, and {Porco}}]{Del_Genio2007}
{Del Genio}, A.~D., {Babara}, J.~M., {Ferrier}, J., {Ingersoll}, A.~P., {West},
  R.~A., {Vasavada}, A.~R., {Spitale}, J., {Porco}, C.~C., 2007. {Saturn eddy
  momentum fluxes and convection: first estimates from Cassini images}. Icarus,
  in press.

\bibitem[{{Del Genio} and {McGrattan}(1990)}]{Del_Genio1990}
{Del Genio}, A.~D., {McGrattan}, K.~B., 1990. {Moist convection and the
  vertical structure and water abundance of Jupiter's atmosphere}. Icarus 84,
  29--53.

\bibitem[{{Dowling}(1995)}]{dowling-1995}
{Dowling}, T.~E., 1995. {Dynamics of Jovian atmospheres}. Ann. Rev. Fluid Mech.
  27, 293--334.

\bibitem[{{Dyudina} et~al.(2007){Dyudina}, {Ingersoll}, {Ewald}, {Porco},
  {Fischer}, {Kurth}, {Desch}, {Del Genio}, {Barbara}, and
  {Ferrier}}]{dyudina-etal-2007}
{Dyudina}, U.~A., {Ingersoll}, A.~P., {Ewald}, S.~P., {Porco}, C.~C.,
  {Fischer}, G., {Kurth}, W., {Desch}, M., {Del Genio}, A., {Barbara}, J.,
  {Ferrier}, J., 2007. {Lightning storms on Saturn observed by Cassini ISS and
  RPWS during 2004 2006}. Icarus 190, 545--555.

\bibitem[{{Dyudina} et~al.(2002){Dyudina}, {Ingersoll}, {Vasavada}, {Ewald},
  and {The Galileo SSI Team}}]{dyudina-etal-2002}
{Dyudina}, U.~A., {Ingersoll}, A.~P., {Vasavada}, A.~R., {Ewald}, S.~P., {The
  Galileo SSI Team}, 2002. {Monte Carlo Radiative Transfer Modeling of
  Lightning Observed in Galileo Images of Jupiter}. Icarus 160, 336--349.

\bibitem[{{Emanuel}(1994)}]{Emanuel_1994}
{Emanuel}, K.~A., 1994. {Atmospheric Convection}. Oxford Univ. Press, New York.

\bibitem[{{Emanuel} and {Raymond}(1993)}]{emanuel-raymond-1993}
{Emanuel}, K.~A., {Raymond}, D., 1993. {The Representation of Cumulus
  Convection in Numerical Models}. American Meteorological Society, Boston.

\bibitem[{{Fegley} et~al.(1991){Fegley}, {Gautier}, {Owen}, and
  {Prinn}}]{fegley-etal-1991}
{Fegley}, B.~J., {Gautier}, D., {Owen}, T., {Prinn}, R.~G., 1991. {Spectroscopy
  and chemistry of the atmosphere of Uranus}. In Uranus (Ed. by J. Bergstrahl,
  University of Arizona Press), 147--203.

\bibitem[{{Flasar} et~al.(2005){Flasar}, {Achterberg}, {Conrath}, {Pearl},
  {Bjoraker}, {Jennings}, {Romani}, {Simon-Miller}, {Kunde}, {Nixon},
  {B{\'e}zard}, {Orton}, {Spilker}, {Spencer}, {Irwin}, {Teanby}, {Owen},
  {Brasunas}, {Segura}, {Carlson}, {Mamoutkine}, {Gierasch}, {Schinder},
  {Showalter}, {Ferrari}, {Barucci}, {Courtin}, {Coustenis}, {Fouchet},
  {Gautier}, {Lellouch}, {Marten}, {Prang{\'e}}, {Strobel}, {Calcutt}, {Read},
  {Taylor}, {Bowles}, {Samuelson}, {Abbas}, {Raulin}, {Ade}, {Edgington},
  {Pilorz}, {Wallis}, and {Wishnow}}]{flasar-etal-2005}
{Flasar}, F.~M., {Achterberg}, R.~K., {Conrath}, B.~J., {Pearl}, J.~C.,
  {Bjoraker}, G.~L., {Jennings}, D.~E., {Romani}, P.~N., {Simon-Miller}, A.~A.,
  {Kunde}, V.~G., {Nixon}, C.~A., {B{\'e}zard}, B., {Orton}, G.~S., {Spilker},
  L.~J., {Spencer}, J.~R., {Irwin}, P.~G.~J., {Teanby}, N.~A., {Owen}, T.~C.,
  {Brasunas}, J., {Segura}, M.~E., {Carlson}, R.~C., {Mamoutkine}, A.,
  {Gierasch}, P.~J., {Schinder}, P.~J., {Showalter}, M.~R., {Ferrari}, C.,
  {Barucci}, A., {Courtin}, R., {Coustenis}, A., {Fouchet}, T., {Gautier}, D.,
  {Lellouch}, E., {Marten}, A., {Prang{\'e}}, R., {Strobel}, D.~F., {Calcutt},
  S.~B., {Read}, P.~L., {Taylor}, F.~W., {Bowles}, N., {Samuelson}, R.~E.,
  {Abbas}, M.~M., {Raulin}, F., {Ade}, P., {Edgington}, S., {Pilorz}, S.,
  {Wallis}, B., {Wishnow}, E.~H., 2005. {Temperatures, Winds, and Composition
  in the Saturnian System}. Science 307, 1247--1251.

\bibitem[{{Gierasch}(1976)}]{1976Icar...29..445G}
{Gierasch}, P.~J., 1976. {Jovian meteorology - Large-scale moist convection}.
  Icarus 29, 445--454.

\bibitem[{{Gierasch} et~al.(2000){Gierasch}, {Ingersoll}, {Banfield}, {Ewald},
  {Helfenstein}, {Simon-Miller}, {Vasavada}, {Breneman}, {Senske}, and {A4
  Galileo Imaging Team}}]{2000Natur.403..628G}
{Gierasch}, P.~J., {Ingersoll}, A.~P., {Banfield}, D., {Ewald}, S.~P.,
  {Helfenstein}, P., {Simon-Miller}, A., {Vasavada}, A., {Breneman}, H.~H.,
  {Senske}, D.~A., {A4 Galileo Imaging Team}, 2000. {Observation of moist
  convection in Jupiter's atmosphere}. \nat 403, 628--630.

\bibitem[{{Glatzmaier} et~al.(2008){Glatzmaier}, {Evonuk}, and
  {Rogers}}]{glatzmaier-etal-2008}
{Glatzmaier}, G.~A., {Evonuk}, M., {Rogers}, T.~M., 2008. {Differential
  rotation in giant planets maintained by density-stratified turbulent
  convection}. ArXiv e-prints 806.

\bibitem[{{Godfrey}(1988)}]{godfrey-1988}
{Godfrey}, D.~A., Nov. 1988. {A hexagonal feature around Saturn's North Pole}.
  Icarus 76, 335--356.

\bibitem[{{Gurnett} et~al.(1990){Gurnett}, {Kurth}, {Cairns}, and
  {Granroth}}]{gurnett-etal-1990}
{Gurnett}, D.~A., {Kurth}, W.~S., {Cairns}, I.~H., {Granroth}, L.~J., 1990.
  {Whistlers in Neptune's magnetosphere - Evidence of atmospheric lightning}.
  J. Geophys. Res. 95, 20967--20976.

\bibitem[{{Hammel} et~al.(2001){Hammel}, {Rages}, {Lockwood}, {Karkoschka}, and
  {de Pater}}]{Hammel2001}
{Hammel}, H.~B., {Rages}, K., {Lockwood}, G.~W., {Karkoschka}, E., {de Pater},
  I., 2001. {New Measurements of the Winds of Uranus}. Icarus 153, 229--235.

\bibitem[{{Heimpel} and {Aurnou}(2007)}]{Heimpel2007}
{Heimpel}, M., {Aurnou}, J., 2007. {Turbulent convection in rapidly rotating
  spherical shells: A model for equatorial and high latitude jets on Jupiter
  and Saturn}. Icarus 187, 540--557.

\bibitem[{{Heimpel} et~al.(2005){Heimpel}, {Aurnou}, and
  {Wicht}}]{2005Natur.438..193H}
{Heimpel}, M., {Aurnou}, J., {Wicht}, J., 2005. {Simulation of equatorial and
  high-latitude jets on Jupiter in a deep convection model}. Nature 438,
  193--196.

\bibitem[{{Huang} and {Robinson}(1998)}]{1998JAtS...55..611H}
{Huang}, H.-P., {Robinson}, W.~A., 1998. {Two-Dimensional Turbulence and
  Persistent Zonal Jets in a Global Barotropic Model.} Journal of Atmospheric
  Sciences 55, 611--632.

\bibitem[{{Hueso} and {S{\'a}nchez-Lavega}(2001)}]{hueso-etal-2001}
{Hueso}, R., {S{\'a}nchez-Lavega}, A., 2001. {A Three-Dimensional Model of
  Moist Convection for the Giant Planets: The Jupiter Case}. Icarus 151,
  257--274.

\bibitem[{{Hueso} et~al.(2002){Hueso}, {S{\'a}nchez-Lavega}, and
  {Guillot}}]{hueso-etal-2002}
{Hueso}, R., {S{\'a}nchez-Lavega}, A., {Guillot}, T., 2002. {A model for
  large-scale convective storms in Jupiter}. Journal of Geophysical Research
  (Planets) 107, 5075--+.

\bibitem[{{Iacono} et~al.(1999){Iacono}, {Struglia}, and
  {Ronchi}}]{1999PhFl...11.1272I}
{Iacono}, R., {Struglia}, M.~V., {Ronchi}, C., 1999. {Spontaneous formation of
  equatorial jets in freely decaying shallow water turbulence}. Physics of
  Fluids 11, 1272--1274.

\bibitem[{{Ingersoll} et~al.(1981){Ingersoll}, {Beebe}, {Mitchell}, {Garneau},
  {Yagi}, and {Muller}}]{ingersoll-etal-1981}
{Ingersoll}, A.~P., {Beebe}, R.~F., {Mitchell}, J.~L., {Garneau}, G.~W.,
  {Yagi}, G.~M., {Muller}, J.-P., 1981. {Interaction of eddies and mean zonal
  flow on Jupiter as inferred from Voyager 1 and 2 images}. J. Geophys. Res.
  86, 8733--8743.

\bibitem[{{Ingersoll} et~al.(2004){Ingersoll}, {Dowling}, {Gierasch}, {Orton},
  {Read}, {S{\'a}nchez-Lavega}, {Showman}, {Simon-Miller}, and
  {Vasavada}}]{ingersoll-etal-2004}
{Ingersoll}, A.~P., {Dowling}, T.~E., {Gierasch}, P.~J., {Orton}, G.~S.,
  {Read}, P.~L., {S{\'a}nchez-Lavega}, A., {Showman}, A.~P., {Simon-Miller},
  A.~A., {Vasavada}, A.~R., 2004. {Dynamics of Jupiter's atmosphere}.
  Jupiter.~The Planet, Satellites and Magnetosphere, pp. 105--128.

\bibitem[{{Ingersoll} et~al.(2000){Ingersoll}, {Gierasch}, {Banfield},
  {Vasavada}, and {A3 Galileo Imaging Team}}]{2000Natur.403..630I}
{Ingersoll}, A.~P., {Gierasch}, P.~J., {Banfield}, D., {Vasavada}, A.~R., {A3
  Galileo Imaging Team}, 2000. {Moist convection as an energy source for the
  large-scale motions in Jupiter's atmosphere}. \nat 403, 630--632.

\bibitem[{{Kaiser} et~al.(1991){Kaiser}, {Desch}, {Farrell}, and
  {Zarka}}]{kaiser-etal-1991}
{Kaiser}, M.~L., {Desch}, M.~D., {Farrell}, W.~M., {Zarka}, P., 1991.
  {Restrictions on the characteristics of Neptunian lightning}. J. Geophys.
  Res. 96, 19043--+.

\bibitem[{{Karoly} et~al.(1998){Karoly}, {Vincent}, and
  {Schrage}}]{karoly-etal-1998}
{Karoly}, D.~J., {Vincent}, D.~G., {Schrage}, J.~M., 1998. {General
  circulation}. Meteorology of the Southern Hemisphere (D.J. Karoly, D.G.
  Vincent, Eds.) American Meteorological Society Monograph 27, 47--85.

\bibitem[{{Li} et~al.(2006){Li}, {Ingersoll}, and {Huang}}]{Li-2006}
{Li}, L., {Ingersoll}, A.~P., {Huang}, X., 2006. {Interaction of moist
  convection with zonal jets on Jupiter and Saturn}. Icarus 180, 113--123.

\bibitem[{{Lian} and {Showman}(2008)}]{lian-showman-2008}
{Lian}, Y., {Showman}, A.~P., 2008. {Deep jets on gas-giant planets}. Icarus
  194, 597--615.

\bibitem[{{Limaye}(1986)}]{1986Icar...65..335L}
{Limaye}, S.~S., 1986. {Jupiter - New estimates of the mean zonal flow at the
  cloud level}. Icarus 65, 335--352.

\bibitem[{{Lindal}(1992)}]{lindal-etal-1992}
{Lindal}, G.~F., 1992. {The atmosphere of Neptune - an analysis of radio
  occultation data acquired with Voyager 2}. Astron. J. 103, 967--982.

\bibitem[{{Lindal} et~al.(1987){Lindal}, {Lyons}, {Sweetnam}, {Eshleman}, and
  {Hinson}}]{lindal-etal-1987}
{Lindal}, G.~F., {Lyons}, J.~R., {Sweetnam}, D.~N., {Eshleman}, V.~R.,
  {Hinson}, D.~P., 1987. {The atmosphere of Uranus - Results of radio
  occultation measurements with Voyager 2}. J. Geophys. Res. 92, 14987--15001.

\bibitem[{{Lindal} et~al.(1985){Lindal}, {Sweetnam}, and
  {Eshleman}}]{lindal-etal-1985}
{Lindal}, G.~F., {Sweetnam}, D.~N., {Eshleman}, V.~R., 1985. {The atmosphere of
  Saturn - an analysis of the Voyager radio occultation measurements}. Astron.
  J. 90, 1136--1146.

\bibitem[{{Lindal} et~al.(1981){Lindal}, {Wood}, {Levy}, {Anderson},
  {Sweetnam}, {Hotz}, {Buckles}, {Holmes}, {Doms}, {Eshleman}, {Tyler}, and
  {Croft}}]{1981JGR....86.8721L}
{Lindal}, G.~F., {Wood}, G.~E., {Levy}, G.~S., {Anderson}, J.~D., {Sweetnam},
  D.~N., {Hotz}, H.~B., {Buckles}, B.~J., {Holmes}, D.~P., {Doms}, P.~E.,
  {Eshleman}, V.~R., {Tyler}, G.~L., {Croft}, T.~A., 1981. {The atmosphere of
  Jupiter - an analysis of the Voyager radio occultation measurements}. J.
  Geophys. Res 86, 8721--8727.

\bibitem[{{Little} et~al.(1999){Little}, {Anger}, {Ingersoll}, {Vasavada},
  {Senske}, {Breneman}, {Borucki}, and {The Galileo SSI Team}}]{Little1999}
{Little}, B., {Anger}, C.~D., {Ingersoll}, A.~P., {Vasavada}, A.~R., {Senske},
  D.~A., {Breneman}, H.~H., {Borucki}, W.~J., {The Galileo SSI Team}, 1999.
  {Galileo Images of Lightning on Jupiter}. Icarus 142, 306--323.

\bibitem[{{Mousis} et~al.(2009){Mousis}, {Marboeuf}, {Lunine}, {Alibert},
  {Fletcher}, {Orton}, {Pauzat}, and {Ellinger}}]{mousis-etal-2009}
{Mousis}, O., {Marboeuf}, U., {Lunine}, J.~I., {Alibert}, Y., {Fletcher},
  L.~N., {Orton}, G.~S., {Pauzat}, F., {Ellinger}, Y., May 2009. {Determination
  of the Minimum Masses of Heavy Elements in the Envelopes of Jupiter and
  Saturn}. Astrophys. J. 696, 1348--1354.

\bibitem[{{Nakajima} et~al.(2000){Nakajima}, {Takehiro}, {Ishiwatari}, and
  {Hayashi}}]{nakajima-etal-2000}
{Nakajima}, K., {Takehiro}, S.-i., {Ishiwatari}, M., {Hayashi}, Y.-Y., Oct.
  2000. {Numerical modeling of Jupiter's moist convection layer}. \grl 27,
  3129--3132.

\bibitem[{{Palotai} and {Dowling}(2008)}]{2008Icar..194..303P}
{Palotai}, C., {Dowling}, T.~E., 2008. {Addition of water and ammonia cloud
  microphysics to the EPIC model}. Icarus 194, 303--326.

\bibitem[{{Porco} et~al.(2005){Porco}, {Baker}, {Barbara}, {Beurle}, {Brahic},
  {Burns}, {Charnoz}, {Cooper}, {Dawson}, {Del Genio}, {Denk}, {Dones},
  {Dyudina}, {Evans}, {Giese}, {Grazier}, {Helfenstein}, {Ingersoll},
  {Jacobson}, {Johnson}, {McEwen}, {Murray}, {Neukum}, {Owen}, {Perry},
  {Roatsch}, {Spitale}, {Squyres}, {Thomas}, {Tiscareno}, {Turtle}, {Vasavada},
  {Veverka}, {Wagner}, and {West}}]{Porco2005}
{Porco}, C.~C., {Baker}, E., {Barbara}, J., {Beurle}, K., {Brahic}, A.,
  {Burns}, J.~A., {Charnoz}, S., {Cooper}, N., {Dawson}, D.~D., {Del Genio},
  A.~D., {Denk}, T., {Dones}, L., {Dyudina}, U., {Evans}, M.~W., {Giese}, B.,
  {Grazier}, K., {Helfenstein}, P., {Ingersoll}, A.~P., {Jacobson}, R.~A.,
  {Johnson}, T.~V., {McEwen}, A., {Murray}, C.~D., {Neukum}, G., {Owen}, W.~M.,
  {Perry}, J., {Roatsch}, T., {Spitale}, J., {Squyres}, S., {Thomas}, P.,
  {Tiscareno}, M., {Turtle}, E., {Vasavada}, A.~R., {Veverka}, J., {Wagner},
  R., {West}, R., 2005. {Cassini Imaging Science: Initial Results on Saturn's
  Atmosphere}. Science 307, 1243--1247.

\bibitem[{{Porco} et~al.(2003){Porco}, {West}, {McEwen}, {Del Genio},
  {Ingersoll}, {Thomas}, {Squyres}, {Dones}, {Murray}, {Johnson}, {Burns},
  {Brahic}, {Neukum}, {Veverka}, {Barbara}, {Denk}, {Evans}, {Ferrier},
  {Geissler}, {Helfenstein}, {Roatsch}, {Throop}, {Tiscareno}, and
  {Vasavada}}]{Porco2003}
{Porco}, C.~C., {West}, R.~A., {McEwen}, A., {Del Genio}, A.~D., {Ingersoll},
  A.~P., {Thomas}, P., {Squyres}, S., {Dones}, L., {Murray}, C.~D., {Johnson},
  T.~V., {Burns}, J.~A., {Brahic}, A., {Neukum}, G., {Veverka}, J., {Barbara},
  J.~M., {Denk}, T., {Evans}, M., {Ferrier}, J.~J., {Geissler}, P.,
  {Helfenstein}, P., {Roatsch}, T., {Throop}, H., {Tiscareno}, M., {Vasavada},
  A.~R., 2003. {Cassini Imaging of Jupiter's Atmosphere, Satellites, and
  Rings}. Science 299, 1541--1547.

\bibitem[{{Read} et~al.(2006){Read}, {Gierasch}, {Conrath}, {Simon-Miller},
  {Fouchet}, and {Yamazaki}}]{Read2006a}
{Read}, P.~L., {Gierasch}, P.~J., {Conrath}, B.~J., {Simon-Miller}, A.,
  {Fouchet}, T., {Yamazaki}, Y.~H., 2006. {Mapping potential-vorticity dynamics
  on Jupiter. I: Zonal-mean circulation from Cassini and Voyager 1 data}.
  Quarterly Journal of the Royal Meteorological Society 132, 1577--1603.

\bibitem[{{Salyk} et~al.(2006){Salyk}, {Ingersoll}, {Lorre}, {Vasavada}, and
  {Del Genio}}]{2006Icar..185..430S}
{Salyk}, C., {Ingersoll}, A.~P., {Lorre}, J., {Vasavada}, A., {Del Genio},
  A.~D., 2006. {Interaction between eddies and mean flow in Jupiter's
  atmosphere: Analysis of Cassini imaging data}. Icarus 185, 430--442.

\bibitem[{{S{\'a}nchez-Lavega} et~al.(2008){S{\'a}nchez-Lavega}, {Orton},
  {Hueso}, {Garc{\'{\i}}a-Melendo}, {P{\'e}rez-Hoyos}, {Simon-Miller}, {Rojas},
  {G{\'o}mez}, {Yanamandra-Fisher}, {Fletcher}, {Joels}, {Kemerer}, {Hora},
  {Karkoschka}, {de Pater}, {Wong}, {Marcus}, {Pinilla-Alonso}, {Carvalho},
  {Go}, {Parker}, {Salway}, {Valimberti}, {Wesley}, and
  {Pujic}}]{sanchez-lavega-etal-2008}
{S{\'a}nchez-Lavega}, A., {Orton}, G.~S., {Hueso}, R., {Garc{\'{\i}}a-Melendo},
  E., {P{\'e}rez-Hoyos}, S., {Simon-Miller}, A., {Rojas}, J.~F., {G{\'o}mez},
  J.~M., {Yanamandra-Fisher}, P., {Fletcher}, L., {Joels}, J., {Kemerer}, J.,
  {Hora}, J., {Karkoschka}, E., {de Pater}, I., {Wong}, M.~H., {Marcus}, P.~S.,
  {Pinilla-Alonso}, N., {Carvalho}, F., {Go}, C., {Parker}, D., {Salway}, M.,
  {Valimberti}, M., {Wesley}, A., {Pujic}, Z., 2008. {Depth of a strong jovian
  jet from a planetary-scale disturbance driven by storms}. \nat 451, 437--440.

\bibitem[{{Schneider} and {Liu}(2009)}]{schneider-liu-2009}
{Schneider}, T., {Liu}, J., 2009. {Formation of Jets and Equatorial
  Superrotation on Jupiter}. Journal of Atmospheric Sciences 66, 579--+.

\bibitem[{{Scott} and {Polvani}(2007)}]{scott-polvani-2007}
{Scott}, R.~K., {Polvani}, L.~M., 2007. {Forced-Dissipative Shallow-Water
  Turbulence on the Sphere and the Atmospheric Circulation of the Giant
  Planets}. Journal of Atmospheric Sciences 64, 3158--+.

\bibitem[{{Scott} and {Polvani}(2008)}]{scott-polvani-2008}
{Scott}, R.~K., {Polvani}, L.~M., Dec. 2008. {Equatorial superrotation in
  shallow atmospheres}. \grl 35, 24202--+.

\bibitem[{{Seiff} et~al.(1998){Seiff}, {Kirk}, {Knight}, {Young}, {Mihalov},
  {Young}, {Milos}, {Schubert}, {Blanchard}, and
  {Atkinson}}]{1998JGR...10322857S}
{Seiff}, A., {Kirk}, D.~B., {Knight}, T.~C.~D., {Young}, R.~E., {Mihalov},
  J.~D., {Young}, L.~A., {Milos}, F.~S., {Schubert}, G., {Blanchard}, R.~C.,
  {Atkinson}, D., 1998. {Thermal structure of Jupiter's atmosphere near the
  edge of a 5-{$\mu$}m hot spot in the north equatorial belt}. J. Geophys. Res
  103, 22857--22890.

\bibitem[{{Shapiro}(1970)}]{Shapiro_1970}
{Shapiro}, R., 1970. {Smoothing, filtering, and boundary effects}. Rev.
  Geophys. Space Phys. 8, 359--387.

\bibitem[{{Showman}(2007)}]{showman-2007}
{Showman}, A.~P., 2007. {Numerical simulations of forced shallow-water
  turbulence: effects of moist convection on the large-scale circulation of
  Jupiter and Saturn.} J. Atmos. Sci 64, 3132--3157.

\bibitem[{{Showman} et~al.(2006){Showman}, {Gierasch}, and
  {Lian}}]{2006Icar..182..513S}
{Showman}, A.~P., {Gierasch}, P.~J., {Lian}, Y., 2006. {Deep zonal winds can
  result from shallow driving in a giant-planet atmosphere}. Icarus 182,
  513--526.

\bibitem[{{Sromovsky} et~al.(1993){Sromovsky}, {Limaye}, and
  {Fry}}]{Sromovsky1993}
{Sromovsky}, L.~A., {Limaye}, S.~S., {Fry}, P.~M., 1993. {Dynamics of Neptune's
  Major Cloud Features}. Icarus 105, 110--141.

\bibitem[{{Sromovsky} et~al.(1983){Sromovsky}, {Revercomb}, {Krauss}, and
  {Suomi}}]{Sromovsky1983}
{Sromovsky}, L.~A., {Revercomb}, H.~E., {Krauss}, R.~J., {Suomi}, V.~E., 1983.
  {Voyager 2 observations of Saturn's northern mid-latitude cloud features -
  Morphology, motions, and evolution}. J. Geophys. Res. 88, 8650--8666.

\bibitem[{{Vasavada} and {Showman}(2005)}]{2005RPPh...68.1935V}
{Vasavada}, A.~R., {Showman}, A.~P., 2005. {Jovian atmospheric dynamics: an
  update after Galileo and Cassini}. Reports of Progress in Physics 68,
  1935--1996.

\bibitem[{{Williams}(1978)}]{1978JAtS...35.1399W}
{Williams}, G.~P., 1978. {Planetary circulations. I - Barotropic representation
  of Jovian and terrestrial turbulence}. Journal of Atmospheric Sciences 35,
  1399--1426.

\bibitem[{{Williams}(2002)}]{Williams_2002}
{Williams}, G.~P., 2002. {Jovian dynamics. Part II: The genesis and
  equilibration of vortex sets}. Journal of Atmospheric Sciences 59,
  1356--1370.

\bibitem[{{Williams}(2003{\natexlab{a}})}]{Williams_2003_3}
{Williams}, G.~P., 2003{\natexlab{a}}. {Barotropic instability and equatorial
  superrotation}. Journal of Atmospheric Sciences 60, 2136--2152.

\bibitem[{{Williams}(2003{\natexlab{b}})}]{Williams_2003_2}
{Williams}, G.~P., 2003{\natexlab{b}}. {Jet sets}. Journal of the
  Meteorological Society of Japan 81, 439--476.

\bibitem[{{Williams}(2003{\natexlab{c}})}]{Williams_2003_1}
{Williams}, G.~P., 2003{\natexlab{c}}. {Jovian Dynamics. Part III. Multiple,
  migrating, and equatorial jets}. Journal of Atmospheric Sciences 60,
  1270--1296.

\bibitem[{{Williams}(2006)}]{Williams_2006}
{Williams}, G.~P., 2006. {Equatorial superrotation and barotropic instability:
  Static stability variants}. Journal of Atmospheric Sciences 63, 1548--1557.

\bibitem[{{Yair} et~al.(1995){Yair}, {Levin}, and {Tzivion}}]{yair-etal-1995}
{Yair}, Y., {Levin}, Z., {Tzivion}, S., 1995. {Lightning generation in a Jovian
  thundercloud: Results from an axisymmetric numerical cloud model.} Icarus
  115, 421--434.

\bibitem[{{Yamazaki} et~al.(2005){Yamazaki}, {Read}, and
  {Skeet}}]{2005P&SS...53..508Y}
{Yamazaki}, Y.~H., {Read}, P.~L., {Skeet}, D.~R., 2005. {Hadley circulations
  and Kelvin wave-driven equatorial jets in the atmospheres of Jupiter and
  Saturn}. \planss 53, 508--525.

\bibitem[{{Zarka} and {Pedersen}(1986)}]{zarka-pedersen-1986}
{Zarka}, P., {Pedersen}, B.~M., 1986. {Radio detection of Uranian lightning by
  Voyager 2}. \nat 323, 605--608.

\end{thebibliography}
\bibliographystyle{elsart-harv}

\clearpage 

\clearpage 

\clearpage 

\clearpage 

\clearpage 

\clearpage 

\clearpage 

\clearpage 

\clearpage 

\clearpage 

\clearpage 

\clearpage 

\clearpage 

\clearpage 

\clearpage 

\clearpage 

\clearpage 

\clearpage 

\clearpage 

\end{document}